\documentclass[pra,twocolumn, showpacs ,superscriptaddress,nofootinbib]{revtex4-1}

\usepackage{makeidx}
\usepackage{multirow}
\usepackage{eurosym} 
\usepackage{amssymb}
\usepackage{amsfonts}
\usepackage{amsmath}
\usepackage{setspace}
\usepackage{bbm}
\usepackage[english]{babel}
\usepackage{graphicx}
\usepackage[applemac]{inputenc}
\usepackage{color}
\usepackage{ifthen}
\usepackage{float}
\usepackage{verbatim}
\usepackage{microtype}

\begin{document}

\newcommand{\leftexp}[2]{{\vphantom{#2}}^{#1}{#2}}
\newcommand{\leftind}[2]{{\vphantom{#2}}_{#1}{#2}}
\newcommand{\bra}[1]{\langle #1|}
\newcommand{\ket}[1]{|#1\rangle}
\newcommand{\braket}[2]{\langle #1|#2\rangle}
\newcommand{\wt}[1]{\widetilde{#1}}
\newcommand{\dblebar}[1]{\overline{\overline{#1}}}
\newcommand{\axe}[1]{\textbf{#1}}
\newcommand{\mat}[1]{\textbf{#1}}
\newcommand{\dvpart}[2]{\frac{\partial #1}{\partial #2}}
\newcommand{\dvtot}[2]{\frac{d #1}{d #2}}
\newcommand{\ddvpart}[2]{\frac{\partial^2 #1}{\partial #2 ^2}}
\newcommand{\ddvtot}[2]{\frac{d^2 #1}{d #2 ^2}}
\newcommand{\dg}{^\dagger}

\def\be{\begin{equation}}
\def\ee{\end{equation}}

\def\bs{\boldsymbol}
\def\bsplit{\begin{split}}
\def\nsplit{\end{split}}
\def\etal{{\it et al.}}
\def\fdsl{\frac{d}{\ell}}
\def\dsl{d/\ell}
\def\ecsl{\left(e^2/\epsilon \ell\right)}
\def\fecsl{\left(\frac{e^2}{\epsilon \ell}\right)}
\def\ua{\uparrow}
\def\da{\downarrow}
\def\cpt{\text{CP}^{3}}
\def\exp{\text{exp}}
\def\dpspn{\frac{2\pi}{\Phi_0}}
\def\vphi{\varphi}
\def\chiT{\chi^{(3)}}

\newcommand{\eq}[1]{\ref{#1})}

\newcommand{\red}{\color[rgb]{0.8,0,0}}
\newcommand{\green}{\color[rgb]{0.0,0.6,0.0}}
\newcommand{\blue}{\color[rgb]{0.0,0.0,0.6}}

\newcommand{\ab}[1]{{\red AB:~#1}} 
\newcommand{\jb}[1]{{\blue JB:~#1}} 

\title{Josephson junction-embedded transmission-line resonators: from Kerr medium to in-line transmon}
\date{\today}

\author{J. Bourassa}
\affiliation{D\'epartement de Physique, Universit\'e de Sherbrooke, Sherbrooke, Qu\'ebec, Canada, J1K 2R1}
\author{F. Beaudoin}
\affiliation{D\'epartement de Physique, Universit\'e de Sherbrooke, Sherbrooke, Qu\'ebec, Canada, J1K 2R1}
\author{Jay M. Gambetta}
\affiliation{IBM T.J. Watson Research Center, Yorktown Heights, NY 10598, USA}
\author{A. Blais}
\affiliation{D\'epartement de Physique, Universit\'e de Sherbrooke, Sherbrooke, Qu\'ebec, Canada, J1K 2R1}

\begin{abstract}
We provide a general method to find the Hamiltonian of a linear circuit in the presence of a nonlinearity. Focussing on the case of a Josephson junction embedded in a transmission-line resonator, we solve for the normal modes of the system by taking into account exactly the effect of the quadratic (i.e.~inductive) part of the Josephson potential. The nonlinearity is then found to lead to self and cross-Kerr effects, as well as beam-splitter type interactions between modes. By adjusting the parameters of the circuit, the Kerr coefficient $K$ can be made to reach values that are weak ($K < \kappa$), strong ($K > \kappa$) or even very strong ($K \gg \kappa$) with respect to the photon-loss rate $\kappa$. In the latter case, the resonator+junction circuit corresponds to an \emph{in-line} version of the transmon. By replacing the single junction by a SQUID, the Kerr coefficient can be tuned in-situ, allowing for example the fast generation of Schrödinger cat states of microwave light. Finally, we explore the maximal strength of qubit-resonator coupling that can be reached in this setting.
\end{abstract}


\maketitle

\section{Introduction}
With their high quality factors and large zero-point electric fields, superconducting transmission-line resonators are versatile tools for the study of quantum mechanical effects in solid-state devices. Resonators have, for example, been used to study the strong coupling regime of cavity QED in electrical circuits~\cite{blais:2004a, wallraff:2004a}, to probe the displacement of a nanomechanical oscillator close to the standard quantum limit~\cite{rocheleau:2010a}, to entangle remote qubits~\cite{sillanpaa:2007a,majer:2007a,dicarlo:2010a,neeley:2010a} and to implement quantum algorithms~\cite{dicarlo:2009a,reed:2012a}. 

In the same way that Josephson junctions are used in qubits for their nonlinearity, these junctions have also been used experimentally to make nonlinear resonators. With their Kerr-type nonlinearity $K(a^\dag a)^2/2$, these resonators have made possible the realization of Josephson bifurcation amplifiers (JBA) to readout the state of superconducting qubits~\cite{boaknin:2007a,metcalfe:2007a,mallet:2009a}, of linear Josephson parametric amplifiers (JPA)~\cite{yurke:1988a,yurke:1989a,yamamoto:2008a,c-b:2007a,c-b:2008a}, of Josephson parametric converter (JPC) between microwave photons of different frequencies~\cite{bergeal:2010a,bergeal:2010b,zakka-bajjani:2011a} and the squeezing of microwave light~\cite{c-b:2008a}. In these experiments, the Kerr nonlinearity is often required to be small with respect to the photon decay rate $\kappa$. For example, the nonlinearity is limiting the dynamic range of  JPAs. Finally, as was theoretically proposed~\cite{bourassa:2009a} and experimentally realized~\cite{niemczyk:2010a}, interrupting a resonator with a Josephson junction can also be used to reach the ultrastrong coupling regime of circuit QED.

In this paper, we give a unified description of this system in a wide range of nonlinearity $K/\kappa$. We first treat the very general problem of finding the normal modes of a continuous linear circuit in which a Josephson junction is embedded. This is done by treating exactly the effect of the quadratic (i.e. inductive) part of the Josephson potential. The non-linearity is then reintroduced and is shown to lead to Kerr nonlinearities and beam-splitter type interactions between modes. By adjusting the parameters of the circuit, the nonlinearity $K$ can be made to reach values that are weak ($K < \kappa$), strong ($K > \kappa$) or even very strong ($K \gg \kappa$) with respect to the photon-loss rate. In the latter case, the resonator+junction circuit corresponds to an \emph{in-line} version of the transmon qubit. By replacing the junction by a SQUID, the Kerr coefficient can be tuned in-situ allowing, for example, the fast generation of Schrödinger cat states of microwave light. In light of these results, we also revisit the question first asked in Ref.~\cite{devoret:2007a} concerning the maximal qubit-resonator coupling strength that is possible in circuit QED. We find that the toy model used there and consisting of lumped LC circuits current biasing a Josephson junction can only be used in a limited range of parameters to predict the coupling.

The paper is organized as follows. In section~\ref{sec:H}, we obtain the normal modes of the transmission-line including the effect of the linearized Josephson junction potential. We then determine the equivalent lumped-element circuit model and reintroduce the Josephson junction's nonlinearity. In section~\ref{sec:quantum-regimes}, the three regimes of $K/\kappa$ mentioned above are explored. Finally, in section~\ref{sec:HowStrong} we discuss the question of the maximal qubit-resonator coupling strength that is possible in circuit QED

\section{Hamiltonian of the nonlinear resonator}
\label{sec:H}

\begin{figure}[tbp]
\centering
\includegraphics[width=0.375\textwidth]{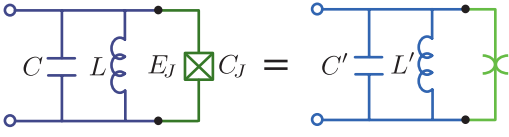}
\caption[]{(Color online) An LC circuit in parallel with a Josephson junction can be equivalently represented by a LC, whose parameters are renormalized by the junction capacitance $C_J$ and the linear Josephson inductance $L_J$, in parallel with a purely nonlinear element represented by the spider-like symbol.}
\label{fig:SimpleLCJ}
\end{figure}

In this section, we solve the general problem of a Josephson junction embedded in a continuous linear circuit. For lumped elements, the approach that we are following can be summarized in a few lines. Indeed, as illustrated in Fig.~\ref{fig:SimpleLCJ}, consider for example a Josephson junction of Josephson energy $E_J$ and capacitance $C_J$ in parallel with a $LC$ oscillator. The Hamiltonian of this circuit is simply
\begin{equation}
H = \frac{q^2}{2C'} +\frac{\delta^2}{2L} - E_J \cos(2\pi\delta/\Phi_0),
\end{equation}
where $C' = C + C_J$. By Taylor expansion of the cosine potential, this can be rewritten in the form
\begin{equation}
H = \frac{q^2}{2C'} + \frac{\delta^2}{2L'} + U_\mathrm{NL}(\delta),
\label{eq:LCSimple}
\end{equation}
with the renormalized inductance $1/L' = (1/L+1/L_J)$ and where we have introduced $1/L_J = (2\pi/\Phi_0)^2E_J$ the Josephson inductance. In this expression, $U_\mathrm{NL}(\delta) = E_J \sum_{n>1} (-1)^{n+1} (2\pi\delta/\Phi_0)^{2n}/(2n)!$ represents the Josephson potential \emph{excluding} the quadratic part $\propto \delta^2$.

As illustrated in Fig.~\ref{fig:SimpleLCJ}, Eq.~\eqref{eq:LCSimple} is simply the Hamiltonian of a linear LC circuit whose parameters $L'$ and $C'$ are renormalized by the presence of the junction and connected in parallel with a purely nonlinear element. Using the notation of Ref.~\cite{manucharyan:2007a}, the latter is represented by a spider-like symbol. Expressing the conjugate operators $q$ and $\delta$ in terms of the oscillator operators $a$ and $a^\dag$, the nonlinear element will contain Kerr-type terms of the form $(a^\dag a)^2$ and whose coefficients can be calculated. We note that, in principle, the nonlinear terms arising from the Josephson potential can be treated to any order as a perturbation on the linear Hamiltonian. This approach can thus be used in any parameter range.

In the rest of this section, this procedure will be followed for a transmission-line resonator in which a Josephson junction is embedded. There, the difficulty comes both from the continuous nature of the system and because it supports multiple modes. We first give the Lagrangian of the resonator including the input/output boundary conditions and the presence of the junction. The normal modes of the Lagrangian where the Josephson potential has been linearized are found. Working in this basis, the Hamiltonian is then obtained. The method presented below is an extension of the work of Wallquist~\etal~\cite{wallquist:2006a} that focused on the first mode of oscillation of a resonator in the presence of a Josephson junction of large Josephson energy (weakly nonlinear regime). The reader not interested in the details of the derivation can jump to section~\ref{sec:ReintroducingNonlinearity} which contains the final form of the quantized Hamiltonian taking into account the nonlinearity.

\subsection{Lagrangian formulation and linearization}

\begin{figure}[tbp]
\centering
\includegraphics[width=0.475\textwidth]{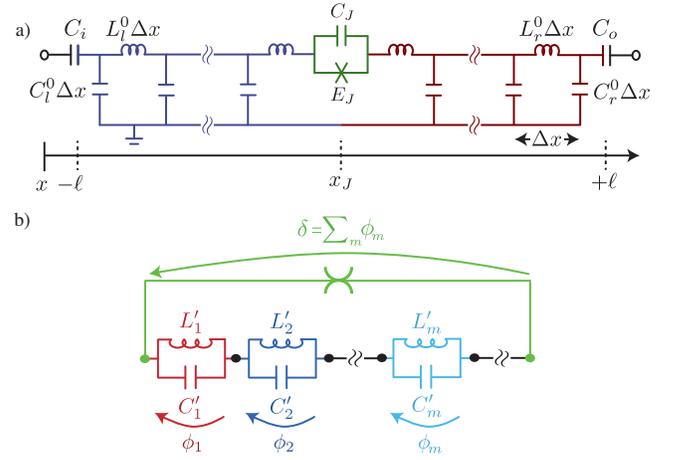}
\caption[]{(Color online) a) Discretized representation of a transmission-line resonator capacitively coupled to input and output ports and with a Josephson junction interrupting the center conductor at position $x_J$. b) Lumped element representation of the transmission line + junction. The normal modes of the resonator, dressed by the junction's capacitance and linear inductance, are represented by LC circuits biasing the junction's nonlinear inductance (spider-like symbol).}
\label{fig:circuits}
\end{figure}

Fig.~\ref{fig:circuits}a) presents schematically a transmission-line resonator with a Josephson junction inserted at position $x_J$ of the center conductor. The resonator, of total length $2 \ell$, is terminated by the input and output capacitors $C_i$ and $C_o$. The Lagrangian of the bare resonator takes the standard form
\be
\mathcal L_r = \int_{-\ell}^\ell \left[ \frac{C^0(x)}{2} \dot \psi^2(x,t) - \frac{(\partial_x \psi(x,t))^2}{2L^0(x)}  \right] dx,
\ee
where, in the flux representation, $\psi(x,t) = \int_{-\infty}^t V(x,t) dt$ where $V(x,t)$ is the voltage, and $C^0(x)$ and $L^0(x)$ are the capacitance and inductance per unit length~\cite{devoret:1997a}. For simplicity, we will take the parameters of the resonator to be piecewise constant. That is, the capacitance $C^0_\mu$ and inductance $L^0_\mu$ per unit length to the left ($\mu = l$) and right ($\mu = r$) of the junction, and the corresponding characteristic impedance $Z_\mu= (L^0_\mu /C^0_\mu)^{1/2}$ and group velocity $v_\mu = (L^0_\mu C^0_\mu)^{-1/2}$, are not assumed to be identical. The generalization to arbitrary $C^0(x)$ and $L^0(x)$ is simple~\cite{bourassa:2009a}.

The input ($\alpha = i$) and output ($\alpha = o$) capacitances $C_\alpha$ are modelled by
\be
\mathcal L_{r\alpha} = \frac{C_\alpha}{2} \left[\dot \psi(x_\alpha,t) - V_\alpha(t) \right]^2.
\ee
In this expression, $V_\alpha(t)$ is the voltage bias at the port $\alpha$ of the resonator and $x_i = -\ell$, $x_o = +\ell$. Finally, the junction is modeled both by its capacitance $C_J$ and Josephson energy $E_J$. In practice, $C_J$ will be a small perturbation on the resonator's total capacitance. Dropping this capacitance would however be akin to dropping the plasma mode of the junction. The contribution of the junction to the Lagrangian is then
\be
\mathcal L_{J} = \frac{C_J}{2} \dot\delta^2 + E_J \cos(2\pi \delta/\Phi_0),
\ee
where $\delta = \psi(x_J^+,t) -  \psi(x_J^-,t)$ is the phase bias of the junction. The junction can be replaced by a SQUID or any other weak link with minimal modification to the theory.

In the same way as for the lumped circuit example above, we expand the cosine potential of the junction such that the total Lagrangian takes the form
\be\label{eq:TotalLagrangian}
\mathcal L =  \mathcal L'_r + \mathcal L_{i} + \mathcal L_{o} - U_\mathrm{NL}(\delta) \equiv \mathcal L_\mathrm{L} - U_\mathrm{NL}(\delta).
\ee
The first term $\mathcal L'_r$ is the resonator Lagrangian including the quadratic contributions of $\mathcal L_J$ and the potential $U_\mathrm{NL}(\delta)$ has been defined as above. We now show how to find the normal modes of the linearized Lagrangian $\mathcal L_\mathrm{L}$.

\subsection{Normal modes decomposition}
\label{sec:normal-modes}
In this section, we are interested in finding the orthogonal basis of normal modes of oscillations of the linear resonator+junction circuit. This is done by solving the Euler-Lagrange equation of motions 
\be
\label{eqn:E-L}
\sum_{\nu=x,t} \partial_\nu \left( \dvpart{\mathcal L_L}{(\partial_\nu \psi(x,t))}\right) - \dvpart{\mathcal L_L}{\psi(x,t)} =0.
\ee
In particular, at the resonator ports $x=\pm \ell$ and the junction's position $x=x_J$, Eq.~\eqref{eqn:E-L} determines the boundary conditions that strongly influence the resonator mode basis.

Away from the junction and the resonator ports, Eq.~\eqref{eqn:E-L} obeys the standard wave equation 
\be
\label{eqn:wave-eq}
\ddot \psi(x,t) =  v_\mu^2 \partial_{xx} \psi(x,t),
\ee
whose solutions are left and right movers with the dispersion relation $\omega_\mu = k_\mu v_\mu$. Since we are looking for modes of the whole resonator, we impose $\omega_l = \omega_r$, or equivalently that the wavevectors obey Snell-Descartes' law $k_r v_r = k_l v_l$. The field $\psi(x,t)$ can be decomposed in terms of these traveling modes  as 
\begin{equation}\label{eq:NormalModeDecomposition}
\psi(x,t) = \sum_m \psi_m(t) u_m(x),
\end{equation}
with $\psi_m(t)$ oscillating at the mode frequency $\omega_m = k_m v_ l$.  In this expression, $k_m$ is the wavevector of the left resonator that we use as a reference and $u_m(x)$ the mode envelope which we now specify using the boundary conditions found from the equation of motion Eq.~\eqref{eqn:E-L}.

Indeed, at $x=\pm \ell$, because of the input and output capacitances, the field must satisfy
\be
\label{eqn:bound-cap}
\begin{split}
\ddot \psi(-\ell,t) - \frac{1}{C_i L^0_l} \partial_x \psi(x,t)|_{x=-\ell} &=  \dot V_i, \\
\ddot \psi(\ell,t) + \frac{1}{C_o L^0_r} \partial_x \psi(x,t)|_{x=\ell} & =  \dot V_o.\\ 
\end{split}
\ee
These equations have homogeneous ($\dot V_{i,o}=0$) and particular ($\dot V_{i,o} \neq 0$) solutions. Moreover, the current on either side of the junction is equal, which imposes
\be
\label{eqn:current-eq}
\begin{split}
\frac{1}{L^0_l} \partial_x \psi(x,t)|_{x=x_J^-} =&\frac{1}{L^0_r} \partial_x \psi(x,t)|_{x=x_J^+} 
= C_J \ddot \delta +  \delta/L_J.
\end{split}
\ee
Without  linearization, the last term of this expression would contain the full Josephson current $I_c \sin(2\pi\delta/\Phi_0)$. 

These constraints can be satisfied by choosing 
\be
\label{eqn:mode-decomp}
u_m(x) = A_m \begin{cases}
				 \sin[k_m(x+\ell) - \varphi_m^i] & -\ell \leq x \leq x_J^-\\
				B_m \sin[k_m'(x-\ell) + \varphi_m^o] & x_J^+ \leq x \leq \ell,
				\end{cases}
\ee
with $k_m' v_r = k_m v_l$. The normalisation contants $A_m$, relative amplitudes $B_m$, phases $\varphi_m^\alpha$ and wavevectors $k_m$ are still to be specified. 

First, $\varphi_m^\alpha$ answers to the homogeneous solution of Eq.~\eqref{eqn:bound-cap}, leading to $\tan\varphi_m^\alpha = \left|Z_\alpha(\omega_m)/Z^0_l\right|$ with $Z_\alpha(\omega)= (i \omega C_\alpha)^{-1}$. Moreover, the first equality of Eq.~\eqref{eqn:current-eq} fixes the relative amplitude $B_m$ of the modes to be
\be
\label{eqn:rel-amps}
B_m = \frac{Z_r^0}{Z_l^0} \frac{\cos[k_m(x_J + \ell) - \varphi_m^i]}{\cos[k_m'(x_J - \ell) + \varphi_m^o]}.
\ee
This corresponds to the impedance mismatch between the two resonator sections at frequency $\omega_m$. Finally, an eigenvalue equation for the wavevector $k_m$ is found by inserting the envelope function Eq.~\eqref{eqn:mode-decomp} in the constraint Eq.~\eqref{eqn:current-eq}. This yields
\be
\label{eqn:e-value}
\bsplit
\left(\frac{Z_r^0}{Z_l^0} \tan\left[k_m\frac{v_l}{v_r}(x_J - \ell) + \varphi_m^o \right] - \tan[k_m(x_J+\ell)- \varphi_m^i]\right) \\
\times \left[ - (k_m \ell)^2  \frac{ C_J }{C^0_l \ell} + \frac{L^0_l \ell}{L_J} \right] = k_m \ell,
\end{split}
\ee
a transcendantal equation whose solutions are found numerically.
\begin{figure}[tbp]
\centering
\includegraphics[width=0.475\textwidth]{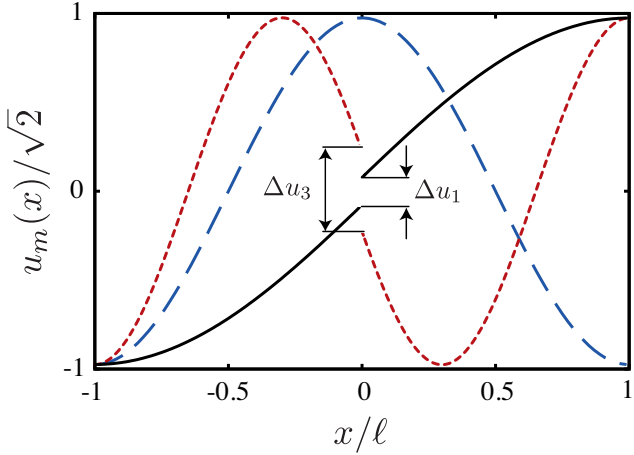}
\caption[]{(Color online) Typical example of the first three normal mode envelopes of a resonator with a Josephson junction at $x_J =0$.}
\label{fig:modes}
\end{figure}

The remaining parameter to be set is the normalization $A_m$. This is done by noting that the $u_m(x)$ obey the inner product~\cite{goldstein:1980a}
\be
\bsplit
\label{eqn:inner-prod1}
\langle u_m \cdot u_n\rangle\equiv &\int_{-\ell} ^\ell  \text d x\, C^0(x) u_m(x) u_n(x) + C_i u_m(-\ell) u_n(-\ell)\\
&+ C_o u_m(\ell) u_n(\ell) + C_J \Delta u_m \Delta u_n \\
=&\,C_\Sigma \delta_{mn}.
\end{split}
\ee
In this expression, $\Delta u_m  = u_m(x_J^+) - u_m(x_J^-)$ is an important parameter corresponding to the mode amplitude difference across the junction. The total capacitance $C_\Sigma~\equiv~\int_{-\ell}^\ell C^0(x) dx + C_i + C_o + C_J$ is here used to fix the normalisation constant $|A_m|$. 

Lastly, from the expressions Eqs.~\eqref{eqn:e-value} and ~\eqref{eqn:inner-prod1}, it is useful to define the inner product of envelope derivatives as they are found to obey a similar orthonormality condition
\be
\bsplit
\label{eqn:inner-prod2}
\langle \partial_x u_m \cdot \partial_x u_n \rangle 
&\equiv \int_{-\ell}^\ell \frac{\text d x}{L^0(x)} \partial_x u_m(x) \partial_x u_n(x) \\
&\qquad + \frac{1}{L_J} \Delta u_m \Delta u_n \\
&=\frac{\delta_{nm}}{L_m} .
\end{split}
\ee
Here, we have defined the mode inductance $L_m^{-1} \equiv C_\Sigma \omega_m^2$ corresponding to the effective inductance of the resonator mode $m$ taking into account the inductance provided by the Josephson junction.

The first three mode envelopes $u_m(x)$ are shown in Fig.~\ref{fig:modes} for a junction symmetrically located at $x_J = 0$. Because they have a finite slope there, the odd modes phase bias the junction which results in a kink $\Delta u_{m,\mathrm{odd}}\neq0$ in the mode envelope. On the other hand, the even modes do not feel the presence of the junction at this location and are unperturbed, $\Delta u_{m, \mathrm{even}}=0$. As first proposed in Ref.~\cite{bourassa:2009a}, this kink can be used to strongly phase bias a qubit and to reach the ultrastrong coupling regime of circuit QED.

\subsection{Hamiltonian of the linearized circuit}

Using the normal mode decomposition Eq.~\eqref{eq:NormalModeDecomposition} and the orthogonality of the mode envelopes Eq.~\eqref{eqn:inner-prod1}, the Lagrangian of the linearized circuit can be expressed as
\begin{equation}
\mathcal L_\mathrm{L}
=
\sum_m \left[ \frac{1}{2} C_\Sigma \dot \psi_m^2 - \frac{\psi_m^2}{2L_m}  
- \dot \psi_m  \sum_{\alpha=i,o} C_\alpha V_\alpha u_m(x_\alpha)\right].
\end{equation}
This immediately leads to the Hamiltonian 
\be
\label{eqn:hamiltonian}
\mathcal H_\mathrm{L} = \sum_m \frac{(q_m - q_{g,m})^2}{2 C_\Sigma} + \frac{\psi_m^2}{2 L_m},
\ee
corresponding to a sum of harmonic oscillators. Here, $q_m = \delta \mathcal L / \delta \dot \psi_m$ is the charge conjugate to $\psi_m$ and $q_{g,m} = \sum_\alpha u_m(x_\alpha) C_\alpha V_\alpha$ is a gate charge associated to the voltage bias at port $\alpha$. 

As will become clear when reintroducing the nonlinearity, for modes having $\Delta u_m \neq 0$, it is advantageous to work with the rescaled conjugate variables $\phi_m = \psi_m \Delta u_m$ and $\rho_m = q_m / \Delta u_m$. In this language, the  above Hamiltonian takes the form 
\be
\label{eqn:h-lumped-Linear}
\mathcal H =\mathcal H_\circ +  \sum_{m\bullet} \left[ \frac{(\rho_m -\rho_{g,m})^2 }{2 C_m'} + \frac{\phi_m^2}{2 L_m' } \right],
\ee
with $C_m' = C_\Sigma/\Delta u_m^2$ and $L_m' = L_m \Delta u_m^2$ and $\rho_{g,m} = q_{g,m}/\Delta u_m$. The symbol $\bullet$ restricts the sum to modes that are affected by the junction and $\circ$ to those that are unaffected by it: $\mathcal H_\circ = \sum_{m\circ} (q_{m}-q_{g,m})^2/(2 C_\Sigma) + \psi_{m}^2/(2L_{m})$. As for the simple lumped-element example presented above, the presence of the junction renormalizes the parameters $C'_m$ and $L'_m$ of the (effective) oscillators. 

To better understand the role played by the junction in the different modes $m$, it is instructive to define the effective resonator capacitor $\wt C_m$ and inductor $\wt L_m$ for mode $m$ (for which $\Delta u_m \neq 0$) in the following way
\be
\bsplit
\wt C_m &\equiv \int_{-\ell}^\ell \text d x \, C^0(x) u_m^2(x) + C_i u^2_m(-\ell) + C_o u^2_m(\ell),\\
1/\wt L_m &\equiv \int_{-\ell}^\ell \frac{\text d x}{L^0(x)} \left[\partial_x u_m(x)\right]^2 .
\end{split}
\ee
To these two quantities are respectively associated the electrostatic and magnetic energy stored only in the resonator and not in the junction. The capacitive and inductive participation ratio of the junction can then be defined as $\eta_{c,m} \equiv C_J /C_m'$ and $\eta_{l,m}\equiv L_m'/L_J$ respectively. As it should, from Eqs.~\eqref{eqn:inner-prod1} and~\eqref{eqn:inner-prod2} the participation ratio of the junction and of the resonator for a given mode $m$ sum to unity
\be
\label{eqn:lumped-elements}
 \eta_{c,m}+\frac{\wt C_m} {C_\Sigma} =1, \qquad \eta_{l,m} +\frac{L_m}{\wt L_m}=1.
\ee
Both ratios are such that $\eta_{c(l),m} \to 0$ in the limit where the junction becomes a short. Moreover, with $C_J \ll C_\Sigma$ in practice, the participation of the junction to the electrostatic energy is small, $\eta_{c,m} \approx 0$.

It is important to understand that the participation ratio of the junction $\eta_{l,m}$ can be quite different from one mode to the other. As mentioned above in relation to Fig.~\ref{fig:modes}, some modes do not feel the presence of the junction. Because of this variation in participation ratio, the mode frequencies are not uniformly spread, $\omega_m \neq m \times  \omega_1$. This inharmonicity can be tuned by changing the position of the junction in the resonator or by changing $E_J$. The latter can be done in-situ by replacing the junction by a SQUID. As will be discussed below, for some applications, this inharmonicity can be advantageous.

\subsection{Reintroducing the nonlinearity}
\label{sec:ReintroducingNonlinearity}

Now that we have the exact modes of the linearized circuits, we reintroduce the nonlinear potential $U_\text{NL}(\delta)$. In terms of the mode decomposition Eq.~\eqref{eq:NormalModeDecomposition}, this takes the form
\be
\label{eqn:potential-nl}
U_\text{NL}(\delta) =  \sum_{i>1} \frac{(-1)^{i+1}}{2i!} \left(\frac{2 \pi}{\Phi_0}\right)^{2i} E_J \left(\sum_m \psi_m \Delta u_m \right)^{2i}.
\ee
Using the rescaled variables defined in Eq.~\eqref{eqn:h-lumped-Linear}, we can get rid of the explicit dependance on $\Delta u_m$ to write the total Hamiltonian in the simple form
\be
\label{eqn:h-lumped}
\mathcal H =\mathcal H_\circ +  \sum_{m\bullet} \left[ \frac{(\rho_m-\rho_{g,m})^2}{2 C_m'} + \frac{\phi_m^2}{2 L_m' } \right]+ U_\text{NL}\left(\sum\nolimits_{m\bullet} \phi_m\right).
\ee
In essence, the system can be modeled by the simple lumped-element circuit presented in Fig.~\ref{fig:circuits}c). This circuit is composed of a discrete set of parallel LC oscillators biasing together a purely nonlinear Josephson inductance. For the remainder of this Article, we will focus only on the modes affected by the junction. 

We now quantize the Hamiltonian and introduce the creation ($a_m^\dagger$) and annihilation ($a_m$) operators of excitations in mode $m$:
\be
\bsplit
\hat \phi_m =& \sqrt{ \frac{\hbar}{2 C_m' \omega_m}} \left(a^\dagger_m + a_m\right), \\
\hat \rho_m =&  i \sqrt{ \frac{\hbar C_m' \omega_m}{2}} \left( a^\dagger_m - a_m\right).
\end{split}
\ee
The above Hamiltonian now takes the form
\be
\hat{\mathcal H} = \hat{\mathcal H}_\mathrm{L} + \hat{\mathcal H}_\mathrm{NL},
\ee
with
\be
\hat{\mathcal H}_\mathrm{L} = \sum_m \hbar \omega_m a^\dagger_m a_m,
\ee
and
\be
\begin{split}
\label{eqn:HNL-photon}
\hat{\mathcal H}_\mathrm{NL} = & 
 \sum_{i>1} \frac{(-1)^{i+1}}{2i!} \left(\frac{2 \pi}{\Phi_0}\right)^{2i} 
\\&
 \times E_J \left[ \sum_{m}\sqrt{ \frac{\hbar}{2 C_m' \omega_m}} (a^\dagger_m + a_m) \right]^{2i}.
\end{split}
\ee

While higher order contributions can easily be taken into account, for simplicity, here we assume small phase fluctuations and consider only the first contribution to the nonlinear term. This yields
\be
\label{eqn:H-4th}
\bsplit
\hat{\mathcal H} \approx
& \sum_m \hbar \omega_m ' a^\dagger_m a_m - \sum_{m,n} \hbar \frac{K_{mn}}{2} a^\dagger_m a_ma^\dagger_n a_n\\
&- \sum_{m\neq n} \hbar \zeta_{mmn} (a^\dagger_n a_m + a^\dagger_m a_n)\\
&- \sum_{l\neq m \neq n}\hbar \zeta_{lmn} \left(a^\dagger_l a_l+ 1/2\right) \left(a^\dagger_m a_n +a^\dagger_n a_m\right),\end{split}
\ee
where we have neglected terms rotating at frequencies faster than $\left|\omega_m - \omega_n\right|$. The nonlinear terms induce a frequency shift $\omega_m' = \omega_m - \sum_n K_{mn}$ on the mode frequencies, as well as Kerr $K_{nn}$, cross-Kerr $K_{mn}$ and beam-splitter like $\zeta_{lmn}$ interactions with amplitudes 
\begin{align}
K_{mm} = & E_{C,m}' \eta_{l,m}/\hbar,\label{eq:SelfKerrCoeff}\\
K_{mn} = & 2 \sqrt{K_{mm} K_{nn}}, \hspace{1cm} \forall~ m\neq n\\
\zeta_{lmn} = & (1-\delta_{lm}/2) \left(K_{ll}^2 K_{mm} K_{nn}\right)^{1/4}.
\end{align}
We express these quantities in terms of the charging energy $E_{C,m}'=e^2/(2C_m')$ and the inductive participation ratio $\eta_{l,m}$. It should not come as a surprise that, in the same way as for the transmon qubit~\cite{koch:2007a}, the self-Kerr coefficient (i.e.~the anharmonicity) is related to the charging energy. Here, however, this nonlinearity is `diluted' by the finite inductance of the transmission line which leads to a non-unity participation ratio $\eta_{l,m}$ of the junction to a given mode $m$. We will come back to the analogies of this system to the transmon below.

Finally, as already pointed out, the Hamiltonian of Eq.~\eqref{eqn:H-4th} includes only the first contribution $\propto\hat\delta^4$ of the nonlinearity. This expansion is thus valid for \emph{weak} nonlinearities with respect to the mode frequencies, $K_{mm}/\omega_m \ll 1$. However, as discussed in the next section, this does not prevent the nonlinearity to be \emph{strong} with respect to the photon damping rate $\kappa$.

\section{Three regimes of nonlinearity}
\label{sec:quantum-regimes}

\begin{figure}[tbp]
\centering
\includegraphics[width=0.475\textwidth]{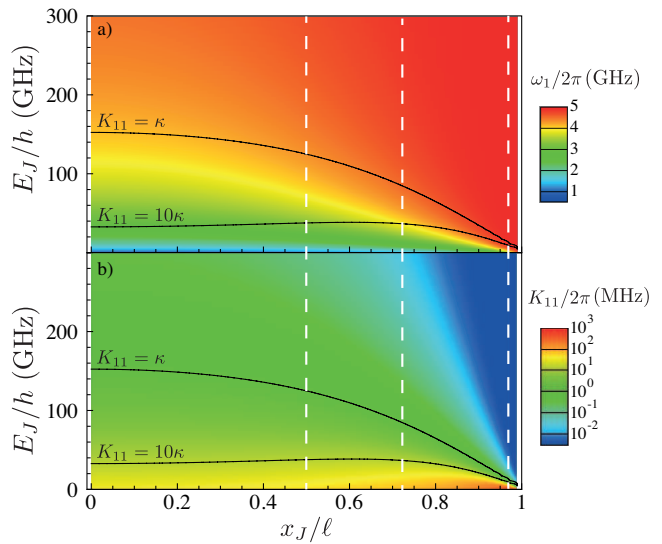}
\caption[]{(Color online) a) Frequency $\omega_1$ and b) Kerr nonlinearity $K_{11}$ of the first mode as a function of $x_J$ and $E_J$. The black lines indicate the contour of constant $K_{11}/\kappa = 1$ and 10, with $\kappa_{\lambda/2}/2\pi = 1.5$~MHz. The dashed white lines indicate $x_J/\ell = 0.5,0.75$ and $0.95$. These positions will be used below. The resonator has a total length $2\ell = 1.2$~cm, characteristic impedance  $Z^0=50~\Omega$ and input and output capacitors $C_i=C_o = 10$~fF. In the absence of the junction, it has a fundamental $\lambda/2$-mode of oscillation at $\omega_{\lambda/2} /2\pi = 4.95$~GHz. Unless otherwise stated, the numerical calculations in this section have been realized with this set of parameters.}
\label{fig:KvsEjXj}
\end{figure}

In this section, we explore analytically and numerically three regimes of nonlinearity. Comparing to the resonator photon-loss rate $\kappa$, we define the weak $K<\kappa$, strong $K>\kappa$ and very strong $K\gg\kappa$ regimes. Here, we will choose the junction location along the length of the resonator to optimize various quantities. 
It is interesting to point out that this choice is not possible with a $\lambda/4$-type resonator where the junction is, by default, at the end of the resonator. The advantage of this additional design flexibility is illustrated in Fig.~\ref{fig:KvsEjXj}. There, we plot the frequency $\omega_1$ and Kerr nonlinearity $K_{11}$ of the first mode of a resonator as a function of junction position $x_J$ and Josephson energy $E_J$. The parameters can be found in the caption of the Figure. By appropriate choice of $x_J$, $E_J$ and, as we will discuss below, the total length $2 \ell$ of the resonator, it is possible to reach the three regimes mentioned above.


While the position can only be chosen at fabrication time, $E_J$ can be tuned in-situ by replacing the single junction with a SQUID with junctions of energy $E_{J1}$ and $E_{J2}$ and using an external flux $\Phi_x$. In this situation, the  Josephson potential $\hat{\mathcal H}_J = - E_J \cos(2\pi \hat\delta/\Phi_0)$ is replaced by~\cite{tinkham:1996a}
\be
\label{eqn:squid-potential}
\begin{split}
\hat{\mathcal H}_J
&= - E_{J\Sigma} \left[\cos\left(\frac{\pi \Phi_x}{\Phi_0}\right) \cos\left(\frac{2\pi \hat\delta}{\Phi_0}\right)
\right.\\
& \qquad\qquad 
+ \left. 
d \sin\left(\frac{\pi \Phi_x}{\Phi_0}\right) \sin\left(\frac{2\pi \hat\delta}{\Phi_0}\right)\right] \\
& = - E_J(\Phi_x) \cos[2\pi(\hat\delta - \delta_0)/\Phi_0],\\
\end{split}
\ee
where
\begin{equation} 
E_J(\Phi_x) = E_{J\Sigma}\cos\left(\frac{\pi \Phi_x}{\Phi_0}\right)\sqrt{1+d^2\tan^2\left(\frac{\pi \Phi_x}{\Phi_0}\right)},
\end{equation}
with $E_{J\Sigma} = E_{j1} + E_{j2}$, $d = |E_{j1} -E_{j2}|/E_{J\Sigma}$ the asymmetry parameter and $\tan \delta_0 = d \tan(\pi \Phi_x/\Phi_0)$. Below, we will drop the phase $\delta_0$ which can be eliminated by a gauge transformation.

Finally, we note the region of large Kerr nonlinearity when the junction is placed close to the of the resonator [bottom right corner of Fig.~\ref{fig:KvsEjXj}b)]. As will be discussed in Sect.~\ref{sec:HowStrong}, in this situation the junction and the small segment of resonator to its right essentially behave as a transmon qubit, the nonlinearity being related to the anharmonicity of the transmon.

\subsection{$K < \kappa$: JBAs, JPAs and JPCs} 
\label{sec:weak}
The regime of weak nonlinearity $K < \kappa$ is realized by working  at large $E_J$ such that the participation ratio $\eta_{l,m}$ of the junction, and in turn the Kerr nonlinearity Eq.~\eqref{eq:SelfKerrCoeff}, is small. This regime has been well studied experimentally, starting by the pioneering experiments in the late 80's of Yurke \etal~with Josephson parametric amplifiers (JPA)~\cite{yurke:1988a,yurke:1989a}. Since, weakly nonlinear resonators have been used as a JPAs~\cite{yamamoto:2008a,c-b:2007a,c-b:2008a}, as Josephson bifurcation amplifier (JBA)~\cite{manucharyan:2007a,metcalfe:2007a,mallet:2009a}, as Josephson parametric converter (JPC)~\cite{bergeal:2010a,bergeal:2010b,zakka-bajjani:2011a} and to squeeze microwave light~\cite{c-b:2008a}. While in this section we are not presenting new ways to exploit this regime, we hope that these results, which can be used to predict the important high-level system parameters (frequencies, nonlinearities, \ldots) from basic device parameters (resonator length, $x_J$, $E_J$, \ldots), will prove useful in practice.

\begin{figure}[tbp]
\centering
\includegraphics[width=0.475\textwidth]{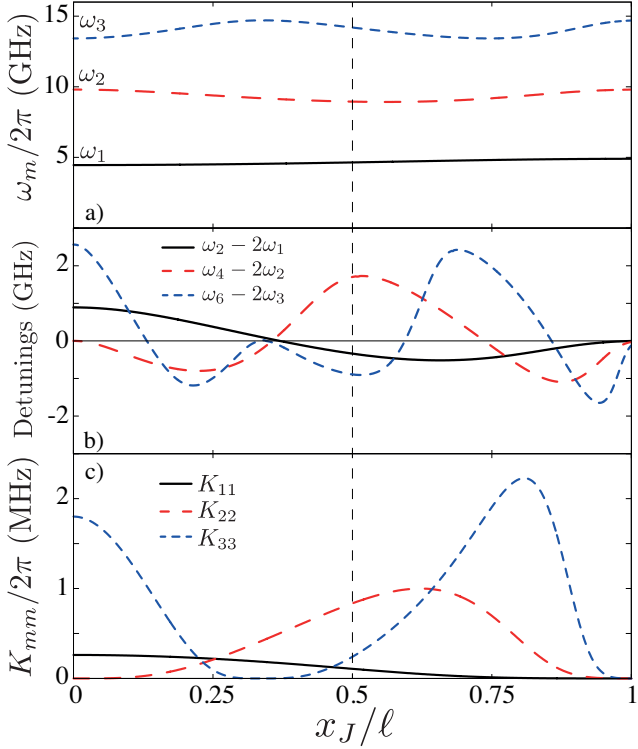}
\caption[]{(Color online) a) Frequencies, b) detunings from two-photon processes and c)  self-Kerr coefficients $K_{mm}$ of the first four normal modes of a nonlinear resonator as a function of the Josephson junction position, starting from the center ($x_J =0$) to the right edge ($x_J = \ell$) of the resonator. Because of the shape of the mode envelopes $u_m(x)$, moving the junction away from the center increases the nonlinearity on the higher resonator modes and thus the inharmonicity. The dashed vertical line indicates at $x_J=\ell/2$ corresponds to the junction location chosen in Fig.~\ref{fig:spectrum-kerr-HiEj-flux}. The parameters are given in the text.}
\label{fig:spectrum-kerr-HiEj}
\end{figure}

We first briefly comment on the use of this device as a JBA, and then move to the JPA and JPC. In the JBA mode, the resonator is driven by a tone of frequency $\omega_d = \omega_m + \Delta$ detuned by $\Delta$ from a given mode $m$. The inharmonicity helps in suppressing population of the other modes, for example if $\left| \omega_m - \omega_{n\neq m} \right| \neq 2 \omega_d$ for the other modes $n$ of the resonator. In this situation, the unwanted modes can be dropped and the Hamiltonian of Eq.~\eqref{eqn:H-4th} takes the simplified form 
\be\label{eq:HJBA}
\bsplit
\mathcal H^{(1)} = \;& \hbar \omega_m' a^\dagger_m a_m - \frac{K_{mm}}{2} (a^\dagger_m a_m)^2 \\
&+ \hbar \left(\epsilon e^{i \omega_d t} a^\dagger_m + \epsilon e^{-i\omega_dt} a_m\right),
\end{split}
\ee
where we have added the drive. Fig.~\ref{fig:spectrum-kerr-HiEj} presents the frequencies, inharmonicity and Kerr amplitude as a function of the junction position (in the range 0 to $\ell$) for a large junction $E_J/h = 636$~GHz. As can be seen in Fig.~\ref{fig:spectrum-kerr-HiEj}b), the detunings to undesirable transitions involving two drive photons can be optimized by moving the junction along the length of the resonator.

We now turn to the JPA and JPC modes of operation. Here, we will assume that a SQUID rather than a single junction is present. Modulating the flux allows for degenerate amplification, as well as non-degenerate amplification and conversion. Indeed, in the presence of a small time-dependant flux $\Phi_\text{rf}(t)$, the Josephson potential Eq.~\eqref{eqn:squid-potential} gains a rf contribution
\be
\begin{split}
\hat{\mathcal H}_J^\mathrm{rf}
&\approx \frac{\varphi_\text{rf}(t)}{2} E_{J\Sigma}\left[ \sin(\varphi_x/2)\cos(2\pi \hat\delta/\Phi_0) \vphantom{\frac{1}{2}}\right.
\\&\qquad
\left.
- d \cos(\varphi_x/2) \sin(2\pi \hat\delta/\Phi_0)\right],
\end{split}
\ee
where we use the notation $\varphi = 2\pi\Phi/\Phi_0$. Taking $\varphi_\text{rf}(t) = \varphi_\text{rf} \cos(\omega_dt)$ this can be rewritten, to second order in $\hat \delta$, as
\be
\bsplit
\hat{\mathcal H}_J^\mathrm{rf}=& - \sum_m  \hbar  (g_m e^{-i \omega_d t} + \text{h.c}) ( a^\dagger_m + a_m)\\
& -  \sum_{m,n} \frac{\hbar}{2} (g_{mn} e^{-i\omega_d t} + \text{h.c})(a^\dagger_m +a _m)(a^\dagger_n + a_n),\\ 
\end{split}
\ee
where the amplitude of the one and two-photon processes are
\begin{align}
g_m=& \left(\dpspn\right)  \frac{d E_{J\Sigma}\cos(\varphi_x/2)}{\sqrt{8 \hbar C_m' \omega_m}}  \varphi_\text{rf} ,\\
g_{mn} = & \left(\dpspn\right)^2 \frac{E_{J\Sigma} \sin(\varphi_x/2)}{4 \sqrt{C_m' C_n' \omega_m \omega_n}}  \varphi_\text{rf} .
\label{eqn:2-photon-drive}
\end{align}
For asymmetric Josephson junctions ($d \neq 0$), both processes are possible. Interestingly, using Eq.~\eqref{eqn:inner-prod2}, it is possible to express $g_{mn}$ in terms of the rate of change of the mode frequencies with respect to the external flux as
\be\label{eq:2-photon-drive-derivatives}
g_{mn} = \frac{1}{\hbar } \sqrt{ \dvpart{ \omega_m}{\Phi_x} \dvpart{\omega_n}{\Phi_x}} \Phi_\mathrm{rf}.
\ee
This relationship, valid for $d\ll 1$ and $\Phi_x \neq \Phi_0/2$, was experimentally verified in Ref.~\cite{zakka-bajjani:2011a}.

We now focus on the symmetric ($d=0$) case in the presence of a non-zero dc component $\varphi_x\neq0$. For $\omega_d \approx \omega_j \pm  \omega_i$, where $i,j$ label two modes of the resonator, the Hamiltonian Eq.~\eqref{eqn:H-4th} including flux driving can be simplified to
\be
\bsplit
\mathcal H^{(2)} =& \sum_{m,n=\{i,j\}} \left[  \hbar \omega_m a^\dagger_m a_m + \frac{ \hbar K_{mn}}{2} a^\dagger_m a_m a^\dagger_na_n \right.\\
& \left.-  \frac{\hbar}{2}(g_{mn} e^{-i\omega_d } + \text{h.c})(a^\dagger_m +a _m)(a^\dagger_n + a_n)\right].
\end{split}
\ee
Because of the repeating indices in the sum, the exchange rate between modes is $g_{mn}$.  For $\omega_d \approx 2 \omega_m$, the last term reduces to $\propto (a^{\dagger\,2}_m + a_m^2)$ corresponding to degenerate parametric amplification~\cite{yamamoto:2008a}. In the non-degenerate mode, with the drive frequency $\omega_d \approx \omega_m + \omega_n$, this term rather reduces to $\propto a^\dagger_m a^\dagger_n + a_m a_n$ which can be used for phase preserving amplification. Finally, for $\omega_d = \omega_m - \omega_n$, we find a beam-splitter like interaction $a^\dagger_m a_n + a_m a^\dagger_n$ between modes. Conversion of microwave photons between two modes in this JPC mode of operation has already been observed~\cite{zakka-bajjani:2011a}. As  noted by these authors, the fidelity of photon frequency conversion will suffer from the nonlinearity $K_{mn}$. Moreover, when operated as a JPA, the nonlinearity will limit the dynamic range of the amplifier and, as discussed in more details in Appendix~\ref{sec:critical_photon_number}, the number of photons that can be present in the resonator before the junction's critical current is reached. The objective is thus to increase the coupling $g_{mn}$ while keeping $K_{mn}$ small. The added design flexibility provided by the choice of the junction's position $x_J$ helps in this regard. As can be seen in Fig.~\ref{fig:KvsEjXj}, the dependance of the mode frequency with respect to $E_J$ (or $\Phi_x$) can be increased by moving the junction along the resonator length without increasing the Kerr non-linearities significantly.

With this in mind, we now compare our theoretical findings to the experimental setting of Ref.~\cite{zakka-bajjani:2011a}. There, a $\lambda/4$ resonator, therefore with the SQUID necessarily located at one end of the resonator, was used to realize the JPC Hamiltonian. The necessary inharmonicity was realized by varying the characteristic impedance of the resonator along its length. In this way, a mode detuning $(\omega_3-\omega_2)-2(\omega_2- \omega_1)= 2 \pi \times 240$ MHz was obtained~\footnote{It is important to note that we take $m=1$ to be the fundamental mode, while the authors of Ref.~\cite{zakka-bajjani:2011a} use $m=0$.}. By biasing the SQUID at $\Phi_x = 0.37 \Phi_0$, the JPC coupling was $g_{12}/2\pi \sim 20$~MHz for a flux modulation amplitude $\Phi_\mathrm{rf}= 0.02 \Phi_0$, while the Kerr coefficients were kept relatively small with $K_{11}/2\pi \sim 0.5$~MHz and $K_{22}/2\pi \sim 4$~MHz.

\begin{figure}[tbp]
\centering
\includegraphics[width=0.475\textwidth]{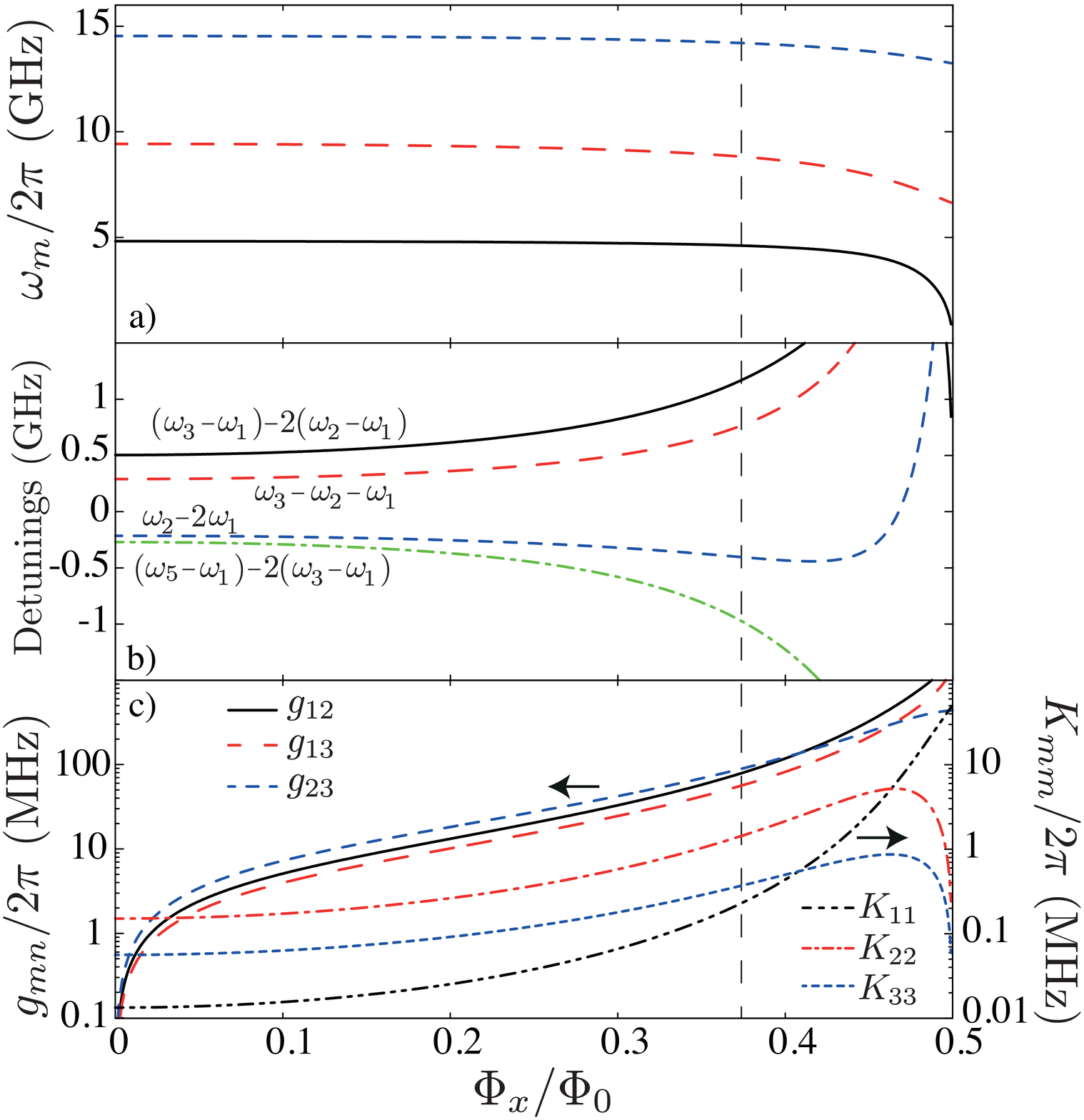}
\caption[]{(Color online) a) Frequencies, b) detunings from high-order photon processes and c) self-Kerr effects $K_{mm}$ and Rabi frequencies $g_{mn}$ for $\Phi_\mathrm{rf} = 0.02 \Phi_0$ of the first three normal modes of a nonlinear resonator with a large Josephson junction, as a function of the external flux $\Phi_x$. The SQUID is placed at position $x_J = \ell/2$. Parameters are given in the text. The vertical dashed line represent at $\Phi_x = 0.37\Phi_0$ corresponds to the operating point of the JPC used in Ref.~\cite{zakka-bajjani:2011a}.}
\label{fig:spectrum-kerr-HiEj-flux}
\end{figure}

Fig.~\ref{fig:spectrum-kerr-HiEj-flux} presents the same parameters (mode frequencies $\omega_m$, inharmonicity, Kerr coefficients $K_{mm}$ and parametric couplings $g_{mn}$) as a function of the external flux $\Phi_x$. In opposition to Ref.~\cite{zakka-bajjani:2011a}, here we consider a $\lambda/2$ resonator with a symmetric SQUID located at $x_J = \ell/2$. This location is also indicated by vertical dashed lines in Figs.~\ref{fig:KvsEjXj} and \ref{fig:spectrum-kerr-HiEj}. We first note that, through most of the $\Phi_x$ range, the detunings to undesired transitions are kept at more than $250$~MHz, similar to Ref.~\cite{zakka-bajjani:2011a}. Moreover, because of the larger flux dependance of the mode frequencies, see Eq.~\eqref{eq:2-photon-drive-derivatives}, for the same small rf amplitude $\Phi_\mathrm{rf} = 0.02 \Phi_0$, we find JPC coupling strengths $\{g_{12},g_{13},g_{23}\}/2\pi \sim \{76,54,86\}$ MHz that are about four times as large. Even with these larger values, the unwanted Kerr nonlinearity remain small at $\{K_{11},K_{22},K_{33}\}/2\pi \sim \{0.21, 1.3 , 0.35\}$ MHz. This increase in coupling strength over nonlinearity should lead to higher fidelities in photon frequency conversion. We note that these parameters have only been manually optimized and a more thorough optimization should lead to better results. Finally, depending on the flux-noise level, it might be more advantageous to work at smaller dc flux bias, where the susceptibility to flux noise is reduced, and increase the rf modulation $\Phi_\mathrm{rf}$ to keep the coupling strength constant.

\subsection{$K > \kappa$: Photon blockade and cat-state generation}

In this section, we will assume that the drive frequency is chosen such that the JPA and JPC terms can be dropped and that a single mode approximation is valid. Similarly to Eq.~\eqref{eq:HJBA}, but dropping the mode index $m$ and the drive, the Hamiltonian reduces to
\be
\label{eqn:h-one-mode}
\hat{\mathcal{H}} = \hbar \omega_r a^\dagger a  - \hbar \frac{K}{2} (a^\dagger a)^2.
\ee
As already illustrated in Fig.~\ref{fig:KvsEjXj} this regime of strong nonlinearity $K > \kappa$ can easily be reached. Moreover, because the Josephson energy is flux-tunable, the Kerr nonlinearity can itself be tuned. Below, we show how to exploit this large nonlinearity, and its tunability, to observe photon blockade~\cite{imamoglu:1997a} and to generate cat states~\cite{walls:1994a}.

\begin{figure}[tbp]
\centering
\includegraphics[width=0.475\textwidth]{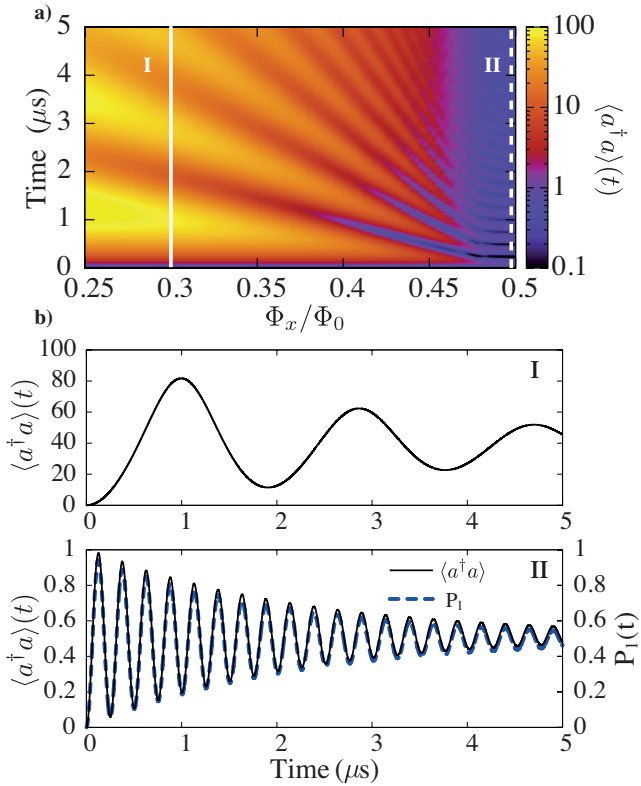}
\caption[]{(Color online) a) Density plot of the average number of photons as a function of time and external flux in presence of an external drive of amplitude $\epsilon/2\pi= 2$~MHz and of frequency $\omega_d = \omega_r- K/2$. b) Line cuts for the values of flux indicated by I and II in panel a) and corresponding respectively to $K/\kappa \sim 1/25$  and $\sim  200$.}
\label{fig:blockade}
\end{figure}

\begin{figure}[tbp]
\centering
\includegraphics[width=0.475\textwidth]{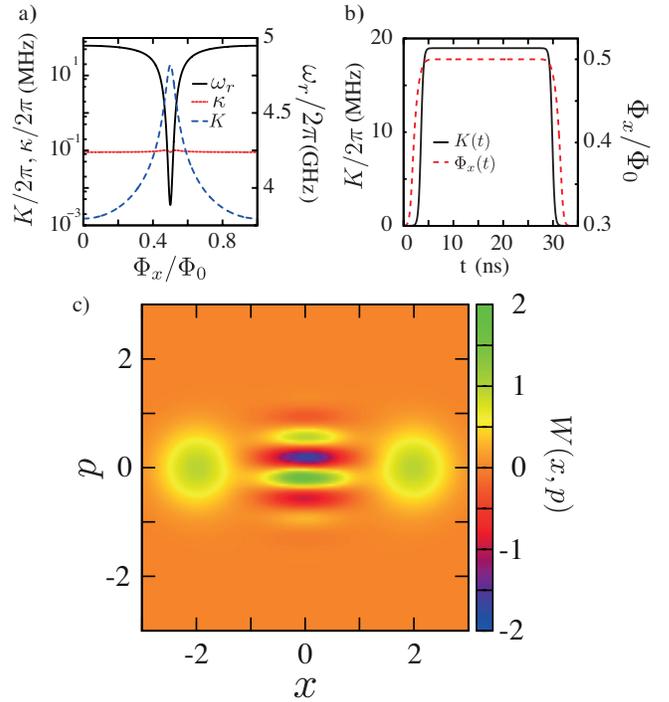}
\caption[]{(Color online) a) Frequency $\omega_r$, radiative relaxation rate $\kappa$ and self-Kerr coefficient $K$ of the first mode as a function of external flux. b) Pulse shape of the external flux and Kerr coefficient as a function of time for Schrödinger cat state generation. c) Density plot of the numerically obtained resonator Wigner function immediately after the flux pulse acting on an initially coherent state $\ket{\alpha =2}$. The device parameters are given in text. }
\label{fig:cat-state}
\end{figure}

For the numerical examples, we will use a SQUID of total Josephson energy $E_{J \Sigma}/2\pi = 622$~GHz interrupting the resonator at $x_J= 3 \ell/4$. These parameter are chosen such that there is a maximum variation of $K$ in a narrow range of flux $\Phi_x$. Indeed, the participation ratio, and hence $K$, change more rapidly with flux near the end of the resonator, as can be seen in Fig.~\ref{fig:KvsEjXj}b). This narrow range should allow for fast-flux tuning to and from operating points with widely different nonlinearities. However, given the dependance of the Kerr effect with the Josephson energy, the extremal points are always near $\Phi_x=0$ and $\Phi_x=\Phi_0/2$, the latter corresponding to the point of largest susceptibility to flux noise. To reduce this susceptibility, we will work with an asymmetric SQUID of asymmetry parameter $d=5\%$. In this case, the point $\Phi_x=\Phi_0/2$ also corresponds to a sweet-spot for flux noise. As can be seen in Fig.~\ref{fig:cat-state}a), with these parameters the Kerr coefficient can be modulated over four orders of magnitude, from $K(\Phi_x =0)/2\pi = 2 \times 10^{-3}$~MHz to $K(\Phi_x=\Phi_0/2)/2\pi = 20$~MHz, over a range of $\Phi_0/2$. At its maximum, the participation ratio of the junction reaches $\eta_{l} = 0.6$.

In order to comfortably reach the strong nonlinear regime $K/\kappa>1$, the coupling capacitances are reduced with respect to the previous section to $C_i=C_o = 2.5$~fF. This corresponds to a very reasonable photon relaxation rate $\kappa/2\pi\approx0.1$ MHz, or equivalently to $T_\kappa=1/\kappa \approx 1.6~\mu$s. With these choices of parameters, the resonator can be brought continuously from the linear regime with $K/\kappa \approx 1/200$ to the strongly nonlinear regime with $K/\kappa \approx 200$ by increasing the flux threading the SQUID loop by half a flux quantum.

In addition to changing the nonlinearity, flux tuning the Josephson energy also changes the mode frequency. This is illustrated by the full black in Fig.~\ref{fig:cat-state}a). Dephasing due to flux noise due to the finite $|\partial \omega /\partial \Phi_x|$ can be evaluated to exceed $10~\mu$s throughout the flux range~\cite{koch:2007a}. In practice, the decoherence time should thus be limited by relaxation with $T_2 \approx 2/\kappa = 3.2~\mu$s. Finally, the red dotted line in Fig.~\ref{fig:cat-state}a) shows a very weak variation of $\kappa$ over the whole flux range. Indeed, while the decay rate from the input (left) port decreases with frequency due to the Ohmic nature of the bath, the reduction is compensated by the enhancement of the decay rate from the output (right) port as the mode amplitude at the port increases with decreasing frequency~\cite{bourassa:2012a}.

Using these parameters, we show in Fig.~\ref{fig:blockade}a) the result of a simulation of the mean photon number $\langle a^\dagger a \rangle (t)$ under irradiation with a continuous tone of amplitude $\epsilon/2\pi = 2$~MHz and frequency $\omega_d(\Phi_x) = \omega_r(\Phi_x) - K(\Phi_x)/2$ as a function of time and for various values of the external flux $\Phi_x$. For small $K/\kappa \sim 1/25$, corresponding to line cut labeled I in panels a) and b), the mean photon number simply exhibits ringing towards its steady-state value. Since the drive is resonant with the $\ket 0 \leftrightarrow \ket 1$ Fock state transition frequency, for a large nonlinearity $K/\kappa \sim  200$, corresponding to the line cut labelled II, the mean photon number rather shows Rabi oscillations with amplitude bounded by $\langle a^\dagger a \rangle=1$. To confirm that these can be interpreted as Rabi oscillations, the probability $P_{1}(t) = \left| \braket{1}{\psi(t)}\right|^2$ is also shown as a blue dashed line in Fig.~\ref{fig:blockade}b). This change of behavior from ringing to Rabi oscillations is also known as photon blockade, where a single photon at a time can enter the resonator because of the large photon-photon interaction $K$~\cite{imamoglu:1997a}. In this regime, the resonator essentially behaves as a qubit with a low anharmonicity $K / \omega_r \sim 0.5\%$. Leakage to higher Fock states can be minimized by pulse shaping~\cite{motzoi:2009a}. We note that photon blockade was already observed in circuit QED using a linear resonator and with a qubit providing the nonlinearity, both in the dispersive~\cite{hoffman:2011a} and resonant~\cite{lang:2011a} regimes.

We now turn to cat-state generation. With the resonator initially prepared in a coherent state $\ket \alpha$, evolution under Hamiltonian Eq.~\eqref{eqn:h-one-mode} for a time $\tau = \pi/K$ will generate a superposition of coherent states with opposite phases~\cite{walls:1994a}
\be
\label{eqn:cat-state}
\ket{\psi_\text{cat}(\alpha)} = \frac{1}{\sqrt{2}} \left( e^{i\pi/4} \ket{ -i\alpha} + e^{-i\pi/4} \ket{i\alpha}\right).
\ee
Here, we suggest to use the tunability of the Kerr coefficient, with an on/off ratio of about $10^4$ for the parameters used here, to prepare with high-fidelity this cat state. 

To prepare this state, we first set the external flux to $\Phi_x=0.3 \Phi_0$ where the resonator is essentially linear. A coherent state $\ket {\alpha}$ in the resonator is prepared using a tone of frequency $\omega_r$ for the appropriate amount of time. As illustrated in Fig.~\ref{fig:cat-state}b), a flux pulse $\Phi_x(t)$ then modulates the nonlinearity $K$ from its small initial value at $\Phi_x=0.3 \Phi_0$ to its maximum at $\Phi_x=0.5\Phi_0$, and back. By choosing the timing such that the total accumulated phase is $\int_0^\tau K(t) dt = \pi$, the field evolves to a cat state. Working with an asymmetric SQUID, the high nonlinearity point where the system spends the most time is a flux sweet-spot, minimizing the effect of flux noise. Moreover, because of the  the large non-linearity $K/2\pi\sim 20$ MHz, the flux excursion is very short with the required phase having been accumulated after a time $\tau = 33$~ns for the present parameters. And as the resonator is back to a small nonlinearity after the protocol, the cat state is preserved (up to phase rotations and damping).

Fig.~\ref{fig:cat-state}c) shows the Wigner function of a cat state prepared using the flux excursion presented in panel b) and with an initial coherent state $\ket{\alpha=2}$. The fidelity $F(\alpha) =  \bra{\psi_\text{cat}(\alpha)}\rho_\mathrm{num}\ket{\psi_\text{cat}(\alpha)}$ of the numerically obtained state $\rho_\mathrm{num}$ to the desired cat state $\ket{\psi_\text{cat}(\alpha)}$ is found to be $F(2)\approx 93.5\%$ for $\alpha =2$ and $F(\sqrt 2) \approx 97\%$ for $\alpha = \sqrt 2$, suggesting that photon loss is the main cause of error.

\subsection{$K \gg \kappa$: in-line transmon}
\label{sec:inline}

\begin{figure}[tbp]
\centering
\includegraphics[width=0.4\textwidth]{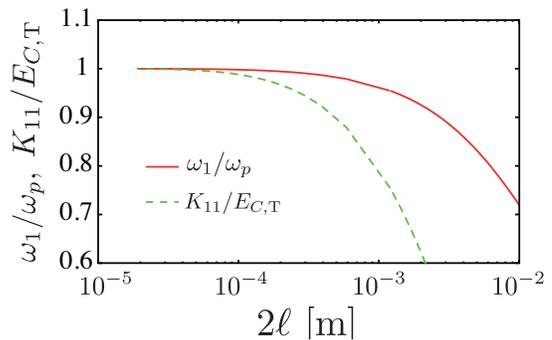}
\caption[]{(Color online) $\omega_1/\omega_p$ (full red line) and $K_{11}/E_{C,T}$ (green dashed line) vs total resonator length $2\ell$. For very short lengths, the circuit essentially behaves as a transmon of frequency $\hbar \omega_p = \sqrt{8 E_{C,T} E_J}$ and anharmonicity $E_{C,\mathrm{T}} = e^2/[2(2\ell C^0/4+C_J)]$.}
\label{fig:KvsLength}
\end{figure}

 By increasing further the participation ratio of the junction, the resonator can be made even more nonlinear. This can easily be achieved by shortening the length of the resonator such that its bare fundamental frequency (in the absence of the junction) lies comfortably above the junction's plasma frequency. In this situation, the first mode of the circuit is essentially the junction's plasma mode, dressed by the resonator. Focussing on this mode, it is useful to write the Hamiltonian in its lumped representation of Eq.~\eqref{eqn:h-lumped}
\be
\label{eqn:h-inline}
\mathcal H \approx  4E_C n ^2 + \frac{E_L}{2} \phi^2 - E_J \cos \phi,
\ee
where $n = q_1/2e$ is the Cooper-pair number operator, conjugate to the phase $\phi$ across the junction. The circuit is described by the charging energy $E_C = (e \Delta u_1)^2/(2 C_\Sigma) $, the linear inductive energy $E_L = (\Phi_0/2\pi)^2/(\wt L_1 \Delta u_1^2)$, while the Josephson energy $E_J$ is unchanged. By taking the one-mode limit and by introducing the inductive energy $E_L$, we have been able to resum the cosine potential in Eq.~\eqref{eqn:h-lumped}. In this effective model $\wt L_1 \Delta u_1^2$ plays the role of the inductance of the dressed plasma mode. 

With the presence of the inductive $\phi^2$ term, Eq.~\eqref{eqn:h-inline} correspond to the Hamiltonian of the fluxonium qubit~\cite{manucharyan:2009b,koch:2009a}. The analogy should however not be pushed too far: the above Hamiltonian is an effective model valid only around the frequency of the mode of interest. Indeed, while in the fluxonium, a large inductance is shunting the junction, there is no such inductive shunt here. As a result, the protection against low-frequency charge noise provided to the fluxonium by the inductance~\cite{koch:2009a} is not present here. In other words, one should be careful about reaching conclusions about low-frequency physics with an effective model valid only around the (relatively large) plasma frequency of the junction.

As a result, despite the presence of the $\phi^2$ term, this system is closer to the transmon~\cite{koch:2007a} than to the fluxonium and, as in Ref.~\cite{devoret:2007a},  we will refer to this qubit  as an \emph{in-line} transmon. As alluded to before, rather than helping, the resonator's inductance dilutes the Josephson inductance, reducing the anharmonicity. Indeed, the transition frequency for the two lowest lying states of the Hamiltonian Eq.~\eqref{eqn:h-inline} can be approximated by $\hbar \omega_{01} \approx \sqrt{8 E_C(E_L+E_J)} $ and the anharmonicity $\alpha=\omega_{21} - \omega_{10}\approx E_C \eta_{l,1}$. In comparison, the transmon transition frequency given by the plasma frequency $\hbar \omega_p=\sqrt{ 8 E_C E_J} $ and it's anharmonicity is given by $\alpha=--E_C$. For a given charging energy, the anharmonicity of the inline transmon is smaller than that of the transmon by the participation ratio $\eta_{l,1}$.

\begin{figure}[tbp]
\centering
\includegraphics[width=0.475\textwidth]{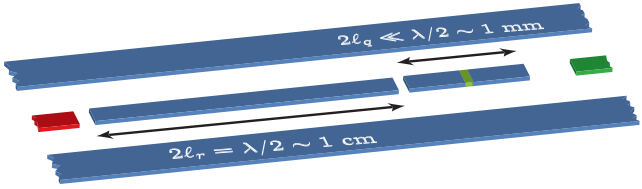}
\caption[]{(Color online) In-line transmon (right) capacitively coupled to a low-Q resonator (left). The in-line transmon can be coherently controlled, for example, using the right (green) control port and the low-Q resonator measured in reflection using the left (red) readout port.}
\label{fig:inlineTransmon}
\end{figure}

As illustrated in Fig.~\ref{fig:KvsLength}, the participation ratio, and thus anharmonicity, is increased by shortening the length of the resonator. In the limit where this length goes to zero, the first mode has a wavevector $k\to0$, corresponding to an envelope that is constant throughout both resonator sections but oscillating out of phase. In this situation, the kink at the junction tends towards $\Delta u_1\to 2$. The participation ratio $\eta_{l,1}$ then tends to unity, such that $E_L \to 0$, $\alpha \to -E_{C,\mathrm{T}}$ and $\omega_1 \to \omega_p$, with $E_{C,\mathrm{T}} = e^2/[2(2\ell C^0/4+C_J)]$ the charging energy of an equivalent transmon~\cite{koch:2007a}. For this purely plasma mode, the in-line transmon reduces to the standard lumped-element transmon. 

In the same way as the transmon, the in-line transmon can be operated in a parameter regime where it is protected against charge noise. However, because of the finite inductance, the protection here improves with the ratio $(E_J+E_L)/E_C$ rather than $E_J/E_C$ as it does for the transmon~\cite{koch:2009a}. In practice however, $E_L \ll E_J$ and the extra factor of $E_L$ should not lead to a significant increase in charge noise protection. However, with respect to the transmon, this qubit could benefit from lower surface losses. Indeed, with its finger capacitor and the associated large electric field, the transmon suffers from surfaces losses~\cite{paik:2011a}. Here, the capacitive shunt is provided by the resonator which does not have (or has less) sharp edges and has a smaller surface area. It is also possible to further decrease the electric field intensity by increasing the gap between the center conductor of the resonator and the ground planes. With quality factors above one million having been achieved with aluminum resonators~\cite{megrant:2012a} and Josephson junctions having been demonstrated to be very coherent~\cite{paik:2011a}, we can expect in-line transmons to have long coherence times.

Finally, in the same way as the transmon, its in-line version can be measured by coupling it to a linear (or non-linear~\cite{mallet:2009a}) resonator of typically lower quality factor. This is schematically illustrated in Fig.~\ref{fig:inlineTransmon}. The readout resonator (left) is capacitively coupled to the short in-line transmon (right). The former can be measured in reflection thought the readout port (left, red) and the latter controlled using the control port (right, green).  A high-Q resonator could be used to mediate entanglement between in-line transmons (not shown)~\cite{blais:2007a}. Finally, we note that, using this setup and since the participation ratio changes with external flux, the presence of the inductive term in Eq.~\eqref{eqn:h-inline} could be observed by spectroscopic measurements of the in-line transmon transition frequencies with respect flux.

\section{How strong can the coupling be?} 
\label{sec:HowStrong}

\begin{figure}
\centering
\includegraphics[width=0.475\textwidth]{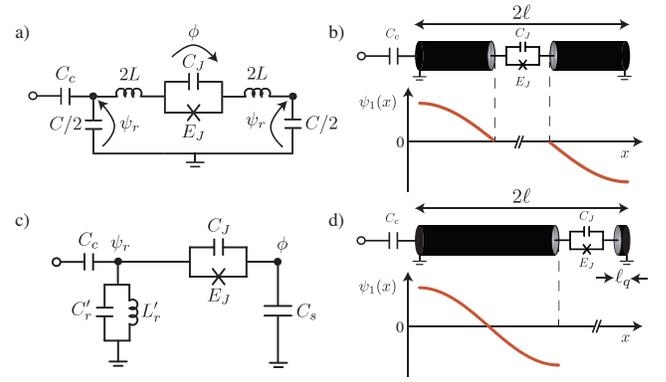}
\caption[]{(Color online) a) Effective circuit of a junction coupled to the current fluctuations of an oscillator as suggested by Devoret \etal~\cite{devoret:2007a}. Close to the resonance and with $C_J \gtrsim C$ such that the $\lambda/2$ mode approximation is valid, this lumped circuit is a good representation of the circuit shown in panel b). c) Effective circuit for a junction coupled to the voltage fluctuations of an oscillator. d) For the physical implementation with a transmission-line resonator to be equivalent to (c) around resonance, it is required that if $\ell_q \ll 2 \ell$ for the $\lambda/2$ mode approximation to be valid. As stated in the text, the effective circuits of panels (a) and (c) are valid representation of the circuits of panels (b) and (d) only around their resonance frequency.}
\label{fig:cqed-circuits}
\end{figure}

The in-line transmon was also suggested in Ref.~\cite{devoret:2007a} as a way to reach the ultrastrong coupling regime of circuit QED, the coupling essentially being between the dressed plasma mode of the junction and a dressed mode of the resonator. In light of the results obtained in this Article, we revisit here this idea. More particularly, we are interested in understanding when the interaction between the plasma mode and a resonator mode  can be approximated by the Rabi-like Hamiltonian 
\be
\label{eqn:rabi}
H_\mathrm{Rabi} =  \hbar \omega_r a^\dagger a + \hbar \omega_{p} b^\dag b  + \frac{\hbar K}{2}  (b^\dag b)^2+ \hbar g (a^\dagger +a)(b^\dagger + b).
\ee
where $a$ is an operator of the dressed resonator mode and $b$ an operator of the dressed plasma mode. In other words, we would like to describe the system using a single resonator mode  and require the plasma mode to preserve a relatively large anharmonicity $K$. In Ref.~\cite{devoret:2007a}, this last requirement was not made and we will therefore arrive at different conclusions here. 

The circuit analyzed in Ref.~\cite{devoret:2007a} is presented in Fig.~\ref{fig:cqed-circuits}a). As illustrated in panel b), in the absence of the junction, this is a lumped element representation of a $\lambda/2$ mode of the resonator with maximums of the voltage at the two ports and a maximum of the current in the center.  In the presence of the junction, in what limit is this lumped-element representation of the continuous circuit valid?

Before answering this question, it is instructive to write the Hamiltonian corresponding to this circuit. Using the conjugate variables $\{\psi_r, \rho_r\}$ and $\{\phi, q\}$ illustrated in Fig.~\ref{fig:cqed-circuits}a), we have
\be
\label{eqn:h-michel}
\begin{split}
H_{\lambda/2} 
&= \frac{\rho_r^2}{2C} + \frac{\left( \psi_r - \phi/2\right)^2}{2L}  + \frac{q^2}{2 C_J} - E_J \cos\left( \frac{2\pi}{\Phi_0} \phi\right)\\
&= 
\left[\frac{\rho_r^2}{2C} + \frac{ \psi_r^2}{2L}\right]
+ \left[ \frac{q^2}{2C_J} + \frac{ \phi^2}{8L} - E_J \cos\left( \frac{2\pi}{\Phi_0} \phi\right)\right]\\
&\quad + \frac{ \psi_r \phi}{2L},
\end{split}
\ee
the three terms corresponding to the resonator, qubit and coupling Hamiltonians respectively. In the same way as in the previous section, we find that the qubit is renormalized by the resonator's inductance and its Hamiltonian takes the form of Eq.~\eqref{eqn:h-inline} with $E_C = e^2/2C_J$ and $E_L = (\Phi_0/2 \pi)^2/4L$.

For $E_J, E_L \gg E_C$, this in-line transmon is well described by a weakly anharmonic oscillator of plasma frequency $ \hbar \omega_p = \sqrt{ 8 E_C(E_J + E_L)}$ and anharmonicity $E_C E_J/(E_J +E_L)$. In this limit, it is useful to introduce the annihilation (creation) operator $b^{(\dag)}$ of the qubit such that  $\phi = \sqrt{\hbar/(2C_J \omega_p)} (b^\dagger +b)$. Also writing $\psi_r = \sqrt{ \hbar/(2 C \omega_r)} (a^\dagger + a)$, the qubit-resonator coupling in Eq.~\eqref{eqn:h-michel} takes the form
\be
\label{eqn:current-coupling}
H_{qr} = \hbar g (a^\dagger + a) (b^\dagger +b).
\ee
As in Ref.~\cite{devoret:2007a}, it is instructive to write the coupling strength $g$ in units of the qubit frequency $\omega_p$:
\be
\label{eqn:g-michel}
\frac{g}{\omega_p} = \frac{\omega_r}{2 \omega_p} \sqrt{ \frac{Z_\mathrm{vac}}{8 \pi \alpha Z_r}} 
\left[ \frac{E_C}{8 (E_J+E_L)}\right]^{1/4},
\ee
where $Z_r = \sqrt{L/C}$ and $\alpha = Z_\mathrm{vac}/(2R_K)$ is the fine structure constant expressed in terms of the vacuum impedance $Z_\mathrm{vac} = 1/\epsilon_0 c \approx 377~\Omega$ and the quantum of resistance $R_K = h/e^2$. Since $Z_r < Z_\mathrm{vac}$, and given the dependence in $1/\sqrt\alpha$, this scheme appears to allow for a coupling ratio $g/\omega_p > 1$ comfortably in the ultrastrong regime. 

We now discuss the constraints on the circuits for the conclusion reached above to be valid. We first note from Eq.~\eqref{eqn:g-michel} that reaching the ultrastrong coupling regime requires $E_J+E_L \ll E_C$~\cite{devoret:2007a}. This is inconsistent with the assumption first made when writing the coupling Hamiltonian in the form of Eq.~\eqref{eqn:current-coupling}. More importantly, the representation of the resonator in Fig.~\ref{fig:cqed-circuits}a), and thus the model Hamiltonian Eq.~\eqref{eqn:h-michel}, is valid only for a $\lambda/2$ mode: we are here assuming a $\lambda/2$ mode \emph{despite} the presence of the junction. Indeed, Eq.~\eqref{eqn:current-coupling} assumes only an inductive coupling between the resonator and the qubit. As we have shown above, the resonator mode is dressed by the junction which then sees a phase drop proportional to $\Delta u_m$. This kink $\Delta u_m$ results in a charge coupling to the resonator mode while simultaneously reducing the inductive coupling Eq.~\eqref{eqn:g-michel}.

This kink can be minimized if the junction capacitance $C_J$ is reasonably large with respect to the resonator capacitance $2\ell C^0$, $C_J \gtrsim 2 \ell C^0$. This minimizes the charge coupling to the advantage of the inductive coupling. This conclusion is also reached when solving numerically the eigenvalue equation Eq.~\eqref{eqn:e-value} requiring $k_1 \approx \pi/(2\ell)$ for $\omega_p > \omega_{\lambda/2}$ [or $k_2 \approx \pi/(2\ell)$ if $\omega_p < \omega_{\lambda/2}$].  For example, satisfying $C_J \gtrsim 2 \ell C^0$ with the  typical resonators parameters given in the beginning of Sect.~\ref{sec:quantum-regimes} requires $C_J \sim 4$~pF. This translates into a very weak charging energy $E_C/h \sim 5$~MHz and correspondingly to a small anharmonicity. For the qubit to be in resonance with the $\lambda/2$ mode at $\omega_p /2\pi = 5$~GHz also requires $(E_J+E_L)/E_C \sim 1.25 \times 10^5$. With these numbers, we find $g/\omega_p \sim 0.2$ from Eq.~\eqref{eqn:g-michel}, a value consistent with numerical calculations. Due to the large junction capacitance involved, it can be difficult in practice to reach the ultrastrong coupling regime with the setup of Fig.~\ref{fig:cqed-circuits}b) and, more generally, realize the Rabi hamiltonian Eq.~\eqref{eqn:rabi} as an increase in the coupling is done at the expense of a reduction of the anharmonicity.

To increase the anharmonicity at roughly constant $g/\omega_p$, a possibility is to slide the qubit away from the center of the $\lambda/2$ mode. In this way, the junction can also be phase biased. As illustrated in Fig.~\ref{fig:cqed-circuits}d), the largest phase bias can be achieved at the end of the $\lambda/2$ mode where the amplitude of the mode envelope is maximal. In the limit where the length of the transmission line $\ell_q$ on one side of the junction is small $\ell_q\ll 2\ell$, the system approximately corresponds to a $\lambda/2$ resonator coupled by the junction to an island of capacitance $C_s=\ell_q C^0$ with negligible inductance. In this setup, the amplitude of the mode envelope at the location of the junction $u(\ell) \approx \sqrt 2$ is only slightly modified by the presence of the junction.

Starting from the total Lagrangian Eq.~\eqref{eq:TotalLagrangian}, neglecting the inductance on the right-hand-side of the junction and focussing on the $\lambda/2$ mode, the effective Hamiltonian of the circuit is found to be 
\be
\label{eqn:h-ours}
H = \frac{\rho_r}{2C_r'} +\frac{\psi_r^2}{2L_r'} + \frac{q^2}{2 C_q} - E_J \cos \left [ \frac{2 \pi}{\Phi_0} ( \phi- \psi_r)\right] - \frac{C_J}{C_q C_r'} \rho_r q,
\ee
where the effective capacitance and inductance of the resonator are $C_r' = (2\ell C^0 + C_J)/2$ and $1/L_r' = C_r' \omega_r$, while the qubit capacitance is $C_q = C_J + C_s$. A lumped element representation of this effective Hamiltonian is shown in Fig.~\ref{fig:cqed-circuits}c). In Eq.~\eqref{eqn:h-ours}, $\psi_r = \sqrt 2 \psi_{1}$ where $\psi_{1}$ is defined in Eq.~\eqref{eq:NormalModeDecomposition}. This Hamiltonian essentially corresponds to a transmon, of charging energy $E_C = e^2/(2C_q)$ and plasma frequency $\hbar \omega_p = \sqrt{ 8 E_C E_J}$, coupled to a LC oscillator through both phase and charge. In the practical limit where $C_J \ll C_s \ll 2 \ell C^0$, the charge interaction is negligible. As stated in Sec.~\ref{sec:inline}, the effective circuit of Fig.~\ref{fig:cqed-circuits}c) is a valid representation of Eq.~\eqref{eqn:h-ours} only around the resonance frequency.

We now evaluate the strength of the dominating coupling. As done above, we work in the limit $E_J/E_C \gg 1$ which here allows us to expand the cosine potential of the junction. Again, by introducing the creation and annihilation of the qubit ($b$) and the resonator ($a$), we find a coupling Hamiltonian of the same form as Eq.~\eqref{eqn:current-coupling} with 
\be
\label{eqn:ratio-ours}
\frac{g}{\omega_p} = \sqrt{\frac{ 2 \pi Z_r' \alpha}{ Z_\mathrm{vac}} } \left( \frac{E_J}{2 E_C}\right)^{1/4},
\ee
where $Z_r' = \sqrt{L_r'/C_r'}= 2 \sqrt{L_r/(2\ell C^0)}$ is the characteristic impedance of the renormalized resonator mode. While Eq.~\eqref{eqn:g-michel} was proportional to $1/\sqrt\alpha$, here the ratio $g/\omega_p$ is proportional to $\sqrt\alpha$. Combined with the small characteristic impedance $Z_r < Z_\mathrm{vac}$, this dependence makes it difficult to reach the ultrastrong coupling regime in this setting. Indeed, with the same resonator parameters as above but now with $E_J/E_C \sim 100$ and $Z_r \sim 15~\Omega$, we find $g/\omega_p\sim 0.15$. However, in contrast to the scheme of Fig.~\ref{fig:cqed-circuits}b), the anharmonicity here is much higher with $E_C/h \sim 300$~MHz.

\begin{figure}[tbp]
\centering
\includegraphics[width=0.45\textwidth]{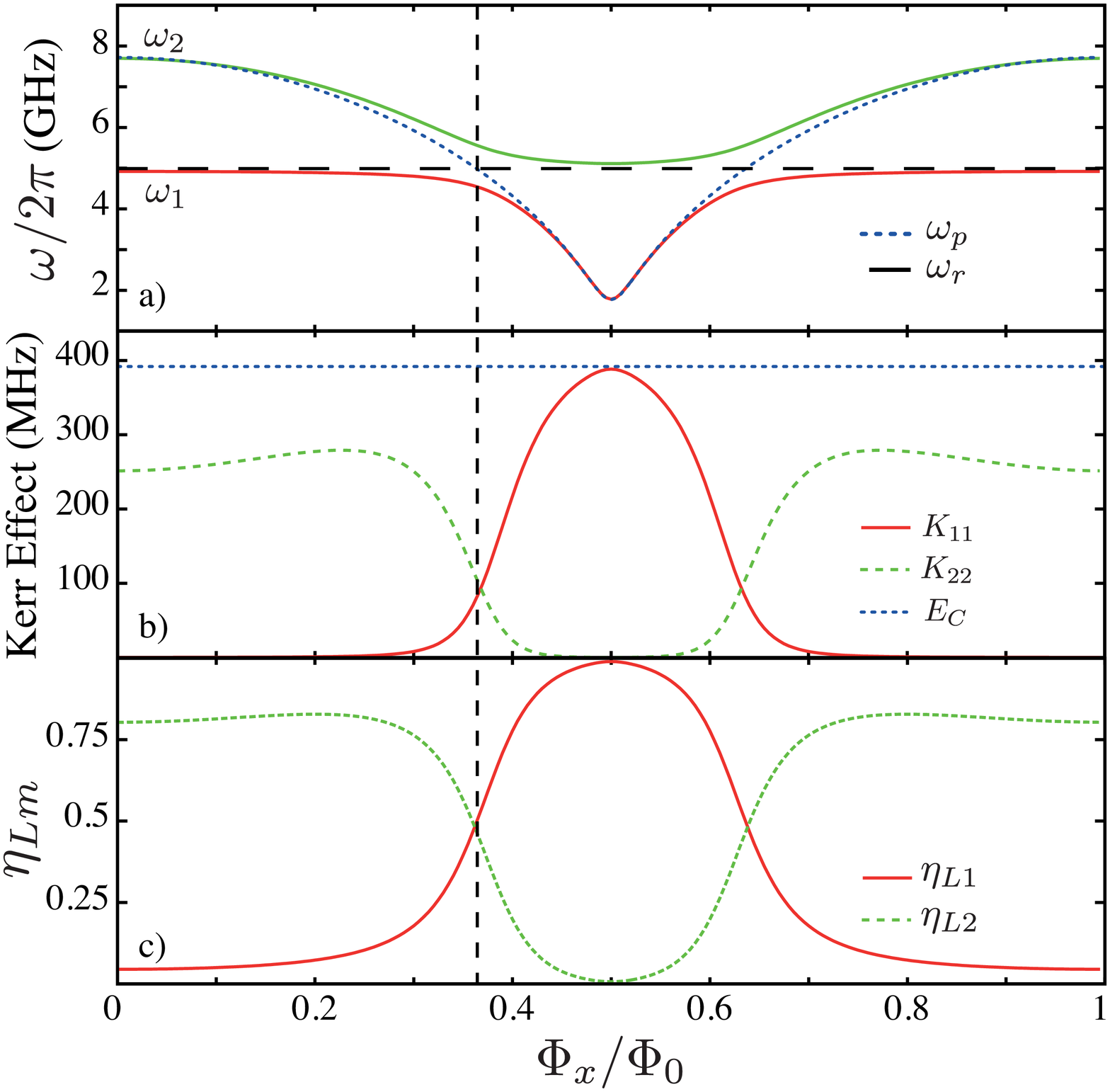}
\caption[]{(Color online) a) Frequencies, b) nonlinear Kerr coefficient and c) participation ratios of the first two normal modes of the resonator for a junction located near the end of the resonator. We have assumed a SQUID of total Josephson energy $E_{J\Sigma}/2\pi = 19$~GHz and asymmetry $d=5\%$ placed at a distance $\ell_q = 260$~$\mu$m from the right extremity of a resonator with bare fundamental frequency $\omega_r/2\pi = 4.95$~GHz. This corresponds to a charging energy of the island of $E_C/2\pi\approx 400$~MHz [blue dotted line in panel b].
}
\label{fig:inline-cqed}
\end{figure}

For completeness, we now compare the effective model of Eq.~\eqref{eqn:h-ours} with numerical simulations of the full system. Fig.~\ref{fig:inline-cqed} presents the normal mode frequencies $\omega_m$, Kerr coefficients $K_{mm}$ and participation ratios $\eta_{l,m}$ as a function of external flux through a SQUID placed near the extremity of a $\lambda/2$ resonator. The parameters can be found in the caption of Fig.~\ref{fig:inline-cqed}. In panel a), the full lines correspond to the normal mode frequencies $\omega_1$ and $\omega_2$, while the dashed line corresponds to $\omega_r$ and the dotted line to $\omega_p$. These last two quantities are evaluated from the effective model Eq.~\eq{eqn:h-ours} and, as expected, agree well with the numerics away from the region of resonance (dashed vertical line). On resonance, an avoided crossing is observed and from which we extract $g/\omega_p = 0.12$, again in good agreement with the effective model.

Simultaneously, both the non-linear Kerr effect and the participation ratio shift from one mode to the other while crossing the resonance, denoting the change of character of the excitations from photon to plasmon (and vice versa). At $\Phi_x = \Phi_0/2$, the participation ratio of the first mode reaches its maximum close to unity while the Kerr nonlinearity reaches $K_{11}\approx E_C /h$ as expected for a transmon. The fact that both $K$ and the participation ratio are not at their maximum value near $\Phi_x=0$ is caused by a residual dressing with the second resonator harmonic close to $10$~GHz (not shown).

These coupling strengths should be contrasted with those obtained using phase-biased flux qubits~\cite{bourassa:2009a} to lumped~\cite{forn-diaz:2010a} and distributed~\cite{niemczyk:2010a} resonators where ratios $g/\omega_q \sim 0.12$ have been reported experimentally. In this setup, it should be possible to comfortably reach even larger coupling ratios while maintaining large anharmonicities~\cite{bourassa:2009a,nataf:2010a}.

\section{Conclusion}

We have presented a general approach to find the normal modes of a linear circuit in which a Josephson junction is embedded. To do so, we included the linear contribution of the junction as a renormalization of the linear circuit parameters. The junction nonlinearity is then reintroduced and leads to Kerr-type nonlinearities and beam-splitter like interactions between modes. This description is most practical for nonlinearities that are weak with respect to the mode frequencies, but can still be large with respect to the photon damping rate. Indeed, we have discussed ways to reach the regimes of weak ($K < \kappa$), strong ($K > \kappa$) and very strong ($K \gg \kappa$) nonlinearity with respect to damping. These results can be used to optimize JBAs, JPAs and JPCs. We have also suggested an approach to generate high-fidelity cat states by tuning rapidly the nonlinearity of the circuit. In the regime of strong nonlinearity, the system behaves as an \emph{in-line} transmon. This qubit could benefit from lower surface losses than the transmon. Finally, we have explored the possibility to reach the ultrastrong coupling regime of circuit QED with the in-line transmon.

We became aware of related works while finishing this paper~\cite{nigg:2012a,leib:2012a}.

\begin{acknowledgments}
We thank M. Boissonneault for help with the numerical simulations, and J. Aumentado, P. Bertet, M. Brink, R. J. Schoelkopf, D. I. Schuster, and particularly M. H. Devoret for fruitful discussions. JB, FB and AB acknowledge funding from FQRNT, NSERC, the Alfred P. Sloan Foundation, and CIFAR. We thank Calcul Qu\'ebec and Compute Canada for computational resources.
\end{acknowledgments}

\appendix
\section{Critical photon number}
\label{sec:critical_photon_number}

A small Josephson energy typically corresponds to large nonlinearities. It however also implies a small critical current $I_c = 2\pi  E_J/\Phi_0$, close to which the junction will switch to the resistive state. Populating the nonlinear resonator with large amplitude fields should thus eventually make the junction switch. This maximal amplitude of the field can be phrased, roughly, in terms of a critical photon number by requiring that $\bra{n_m} \hat{\mathcal{H}}_L \ket{n_m} < E_J$, where $\ket{n_m}$ is a Fock state of mode $m$. A lower bound for this photon occupation can then be expressed as 
\be
n^c_m > \frac{E_J} {\hbar \omega_m}  = \sqrt{ \eta_{l,m}} \sqrt{\frac{E_J}{8E_{C,m}'}} = \eta_{l,m} \sqrt{ \frac{E_J}{8\hbar K_{mm}}}.
\ee
As it should, this result is similar to what is found for the number of levels $\sim  \sqrt{E_J/8E_C}$ in the potential well of a transmon qubit. The presence of $\eta_{l,m}$ reflecting here the reduced participation ratio of the junction. Of course, for transmons, exceeding the critical photon number does not lead to switching because the junction is voltage and not current biased. 

For the parameter regimes studied here, where the ratio $E_J/E_{C,m}' \sim 10^2-10^5$, the number of accessible states is always much larger than unity with $n^c_1 \sim 10$ in the strong nonlinear regime and $n^c_m \gtrsim 10^2$ in the weakly nonlinear regime.

\singlespacing


\begin{thebibliography}{10}
\newcommand{\enquote}[1]{``#1''}





\bibitem{blais:2004a}
A.~Blais, R.-S. Huang, A.~Wallraff, S.~M. Girvin, and R.~J. Schoelkopf,
  Physical Review A \textbf{69},
  062320 (2004).

\bibitem{wallraff:2004a}
A.~Wallraff, D.~I. Schuster, A.~Blais, L.~Frunzio, R.~S. Huang, J.~Majer,
  S.~Kumar, S.~M. Girvin, and R.~J. Schoelkopf, Nature \textbf{431}, 162
  (2004).

\bibitem{rocheleau:2010a}
T.~Rocheleau, T.~Ndukum, C.~Macklin, J.~B. Hertzberg, A.~A. Clerk, and K.~C.
  Schwab, Nature \textbf{463}, 72 (2010).

\bibitem{sillanpaa:2007a}
M.~A. Sillanpaa, J.~I. Park, and R.~W. Simmonds, Nature \textbf{449}, 438
  (2007).

\bibitem{majer:2007a}
J.~Majer, J.~M. Chow, J.~M. Gambetta, J.~Koch, B.~R. Johnson, J.~A. Schreier,
  L.~Frunzio, D.~I. Schuster, A.~A. Houck, A.~Wallraff, A.~Blais, M.~H.
  Devoret, S.~M. Girvin, and R.~J. Schoelkopf, Nature \textbf{449}, 443 (2007).

\bibitem{dicarlo:2010a}
L.~DiCarlo, M.~D. Reed, L.~Sun, B.~R. Johnson, J.~M. Chow, J.~M. Gambetta,
  L.~Frunzio, S.~M. Girvin, M.~H. Devoret, and R.~J. Schoelkopf, Nature
  \textbf{467}, 574 (2010).

\bibitem{neeley:2010a}
M.~Neeley, R.~C. Bialczak, M.~Lenander, E.~Lucero, M.~Mariantoni, A.~D.
  O'Connell, D.~Sank, H.~Wang, M.~Weides, J.~Wenner, Y.~Yin, T.~Yamamoto, A.~N.
  Cleland, and J.~M. Martinis, Nature \textbf{467}, 570 (2010).

\bibitem{dicarlo:2009a}
L.~DiCarlo, J.~M. Chow, J.~M. Gambetta, L.~S. Bishop, B.~R. Johnson, D.~I.
  Schuster, J.~Majer, A.~Blais, L.~Frunzio, S.~M. Girvin, and R.~J. Schoelkopf,
  Nature \textbf{460}, 240 (2009).

\bibitem{reed:2012a}
M.~D. Reed, L.~DiCarlo, S.~E. Nigg, L.~Sun, L.~Frunzio, S.~M. Girvin, and R.~J.
  Schoelkopf, Nature \textbf{482}, 382 (2012).

\bibitem{boaknin:2007a}
E.~Boaknin, V.~E. Manucharyan, S.~Fissette, M.~Metcalfe, L.~Frunzio, R.~Vijay,
  I.~Siddiqi, A.~Wallraff, R.~J. Schoelkopf, and M.~Devoret, arXiv
  \textbf{cond-mat} (2007).

\bibitem{metcalfe:2007a}
M.~Metcalfe, E.~Boaknin, V.~Manucharyan, R.~Vijay, I.~Siddiqi, C.~Rigetti,
  L.~Frunzio, R.~J. Schoelkopf, and M.~H. Devoret, Phys. Rev. B \textbf{76},
  174516 (2007).

\bibitem{mallet:2009a}
F.~Mallet, F.~R. Ong, A.~Palacios-Laloy, F.~Nguyen, P.~Bertet, D.~Vion, and
  D.~Esteve, Nat Phys \textbf{5}, 791 (2009).

\bibitem{yurke:1988a}
B.~Yurke, P.~G. Kaminsky, R.~E. Miller, E.~A. Whittaker, A.~D. Smith, A.~H.
  Silver, and R.~W. Simon, Phys. Rev. Lett. \textbf{60}, 764 (1988).

\bibitem{yurke:1989a}
B.~Yurke, L.~R. Corruccini, P.~G. Kaminsky, L.~W. Rupp, A.~D. Smith, A.~H.
  Silver, R.~W. Simon, and E.~A. Whittaker, Phys. Rev. A \textbf{39}, 2519
  (1989).

\bibitem{yamamoto:2008a}
T.~Yamamoto, K.~Inomata, M.~Watanabe, K.~Matsuba, T.~Miyazaki, W.~D. Oliver,
  Y.~Nakamura, and J.~S. Tsai, Applied Physics Letters \textbf{93}, 042510
  (2008).

\bibitem{c-b:2007a}
M.~A. Castellanos-Beltran and K.~W. Lehnert, Applied Physics Letters
  \textbf{91}, 083509 (2007).

\bibitem{c-b:2008a}
M.~A. Castellanos-Beltran, K.~D. Irwin, G.~C. Hilton, L.~R. Vale, and K.~W.
  Lehnert, Nat Phys \textbf{4}, 929 (2008).

\bibitem{bergeal:2010a}
N.~Bergeal, R.~Vijay, V.~E. Manucharyan, I.~Siddiqi, R.~J. Schoelkopf, S.~M.
  Girvin, and M.~H. Devoret, Nat Phys \textbf{6}, 296 (2010).

\bibitem{bergeal:2010b}
N.~Bergeal, F.~Schackert, M.~Metcalfe, R.~Vijay, V.~E. Manucharyan, L.~Frunzio,
  D.~E. Prober, R.~J. Schoelkopf, S.~M. Girvin, and M.~H. Devoret, Nature
  \textbf{465}, 64 (2010).

\bibitem{zakka-bajjani:2011a}
E.~Zakka-Bajjani, F.~Nguyen, M.~Lee, L.~R. Vale, R.~W. Simmonds, and
  J.~Aumentado, Nat Phys \textbf{7}, 599 (2011).

\bibitem{bourassa:2009a}
J.~Bourassa, J.~M. Gambetta, J.~A.~A.~Abdumalikov, O.~Astafiev, Y.~Nakamura,
  and A.~Blais, Physical Review A \textbf{80}, 032109 (2009).

\bibitem{niemczyk:2010a}
T.~Niemczyk, F.~Deppe, H.~Huebl, E.~P. Menzel, F.~Hocke, M.~J. Schwarz, J.~J.
  Garcia-Ripoll, D.~Zueco, T.~Hummer, E.~Solano, A.~Marx, and R.~Gross, Nat
  Phys \textbf{6}, 772 (2010).

\bibitem{devoret:2007a}
M.~Devoret, S.~Girvin, and R.~Schoelkopf, Annalen der Physik \textbf{16}, 767
  (2007).

\bibitem{manucharyan:2007a}
V.~E. Manucharyan, E.~Boaknin, M.~Metcalfe, R.~Vijay, I.~Siddiqi, and
  M.~Devoret, Physical Review B \textbf{76}, 014524 (2007).

\bibitem{wallquist:2006a}
M.~Wallquist, V.~S. Shumeiko, and G.~Wendin, Phys. Rev. B \textbf{74}, 224506
  (2006).

\bibitem{devoret:1997a}
M.~H. Devoret, in \emph{Les Houches, Session LXIII, 1995}, edited by
  S.~Reynaud, E.~Giacobino, and J.~Zinn-Justin (Elsevier Science, Amsterdam,
  1997), p. 351.

\bibitem{goldstein:1980a}
H.~Goldstein, \emph{Classical Mechanics} (Addison Wesley, 1980).

\bibitem{koch:2007a}
J.~Koch, T.~M. Yu, J.~Gambetta, A.~A. Houck, D.~I. Schuster, J.~Majer,
  A.~Blais, M.~H. Devoret, S.~M. Girvin, and R.~J. Schoelkopf, Physical Review
  A \textbf{76}, 042319 (2007).

\bibitem{tinkham:1996a}
M.~Tinkham, \emph{Introduction to Superconductivity} (Dover, 1996), second
  edition ed.

\bibitem{imamoglu:1997a}
A.~Imamoglu, A., H.~Schmidt, G.~Woods, and M.~Deutsch, Phys. Rev. Lett.
  \textbf{79}, 1467 (1997).

\bibitem{walls:1994a}
D.~Walls and G.~Milburn, \emph{Quantum optics} (Spinger-Verlag, Berlin, 1994).

\bibitem{bourassa:2012a}
J.~Bourassa, J.~M. Gambetta, and A.~Blais, \enquote{In preparation},  (2012).

\bibitem{motzoi:2009a}
F.~Motzoi, J.~M. Gambetta, P.~Rebentrost, and F.~K. Wilhelm, Phys. Rev. Lett.
  \textbf{103}, 110501 (2009).

\bibitem{hoffman:2011a}
A.~J. Hoffman, S.~J. Srinivasan, S.~Schmidt, L.~Spietz, J.~Aumentado, H.~E.
  T\"ureci, and A.~A. Houck, Phys. Rev. Lett. \textbf{107}, 053602 (2011).

\bibitem{lang:2011a}
C.~Lang, D.~Bozyigit, C.~Eichler, L.~Steffen, J.~M. Fink, A.~A. Abdumalikov,
  M.~Baur, S.~Filipp, M.~P. da~Silva, A.~Blais, and A.~Wallraff, Phys. Rev.
  Lett. \textbf{106}, 243601 (2011).

\bibitem{manucharyan:2009b}
V.~E. Manucharyan, J.~Koch, L.~I. Glazman, and M.~H. Devoret, Science
  \textbf{326}, 113 (2009).

\bibitem{koch:2009a}
J.~Koch, V.~Manucharyan, M.~H. Devoret, and L.~I. Glazman, Phys. Rev. Lett.
  \textbf{103}, 217004 (2009).

\bibitem{paik:2011a}
H.~Paik, D.~I. Schuster, L.~S. Bishop, G.~Kirchmair, G.~Catelani, A.~P. Sears,
  B.~R. Johnson, M.~J. Reagor, L.~Frunzio, L.~I. Glazman, S.~M. Girvin, M.~H.
  Devoret, and R.~J. Schoelkopf, Phys. Rev. Lett. \textbf{107}, 240501 (2011).

\bibitem{megrant:2012a}
A.~Megrant, C.~Neill, R.~Barends, B.~Chiaro, Y.~Chen, L.~Feigl, J.~Kelly,
  E.~Lucero, M.~Mariantoni, P.~J.~J. O'Malley, D.~Sank, A.~Vainsencher,
  J.~Wenner, T.~C. White, Y.~Yin, J.~Zhao, C.~J. Palmstr{\o}m, J.~M. Martinis,
  and A.~N. Cleland, Applied Physics Letters \textbf{100}, 1135210 (2012).

\bibitem{blais:2007a}
A.~Blais, J.~Gambetta, A.~Wallraff, D.~I. Schuster, S.~M. Girvin, M.~H.
  Devoret, and R.~J. Schoelkopf, Physical Review A \textbf{75}, 032329 (2007).

\bibitem{forn-diaz:2010a}
P.~Forn-D\'iaz, J.~Lisenfeld, D.~Marcos, J.~J. Garc\'ia-Ripoll, E.~Solano,
  C.~J. P.~M. Harmans, and J.~E. Mooij, Phys. Rev. Lett. \textbf{105}, 237001
  (2010).

\bibitem{nataf:2010a}
P.~Nataf and C.~Ciuti, Phys. Rev. Lett. \textbf{104}, 023601 (2010).

\bibitem{nigg:2012a}
S.~E. Nigg, H.~Paik, B.~Vlastakis, G.~Kirchmair, S.~Shankar, L.~Frunzio, M.~H.
  Devoret, R.~J. Schoelkopf, and S.~M. Girvin, Phys. Rev. Lett. \textbf{108},
  240502 (2012).

\bibitem{leib:2012a}
M.~{Leib}, F.~{Deppe}, A.~{Marx}, R.~{Gross}, and M.~{Hartmann}, ArXiv:1202.3240
   (2012).

\end{thebibliography}
\end{document}